# Network Refinement: A unified framework for enhancing signal or removing noise of networks

Jiating Yu, Jiacheng Leng, and Ling-Yun Wu*

*Abstract*—**Networks are widely used in many fields for their powerful ability to provide vivid representations of relationships between variables. However, many of them may be corrupted by experimental noise or inappropriate network inference methods that inherently hamper the efficacy of network-based downstream analysis. Consequently, it's necessary to develop systematic methods for denoising networks, namely, improve the Signal-to-Noise Ratio (SNR) of noisy networks. In this paper, we have explored the properties of network signal and noise and proposed a novel network denoising framework called Network Refinement (NR) that adjusts the edge weights by applying a nonlinear graph operator based on a diffusion process defined by random walk on the graph. Specifically, this unified framework consists of two closely linked approaches named NR-F and NR-B, which improve the SNR of noisy input networks from two different perspectives: NR-F aims at enhancing signal strength, while NR-B aims at weakening noise strength. Users can choose from which angle to improve the SNR of the network according to the characteristics of the network itself. We show that NR can significantly refine the quality of many networks by several applications on simulated networks and typical real-world biological and social networks.**

## I. INTRODUCTION

NETWORKS are ubiquitous in various fields including social networks such as Facebook networks, co-authorship collaborations networks, and biological networks such as gene regulatory networks (GRN), protein-protein interaction (PPI) networks [1]-[3], [10], etc. The weighted edges between pairs of nodes provide an efficient representation for variable interdependencies, with the edge weight typically corresponding to the confidence or the strength of a given relationship [1]. However, many networks are noisy due to the limitations of measurement technology, natural variations of information, and the inherent flaw of the network inference methods, which hamper the discovery of network properties and characteristics [3]. Thus, it's valuable to develop methods for enhancing the Signal-to-Noise Ratio (SNR) of noisy networks.

This work has been supported by the National Key Research and Development Program of China (No. 2020YFA0712402) and the National Natural Science Foundation of China (No. 11631014). (*Corresponding author: Ling-Yun Wu.*)

The authors are with IAM, MADIS, NCMIS, Academy of Mathematics and Systems Science, Chinese Academy of Sciences, Beijing 100190, China and School of Mathematical Sciences, University of Chinese Academy of Sciences, Beijing 100049, China (email: yujiating@amss.ac.cn; jcleng@amss.ac.cn; lywu@amss.ac.cn)

To overcome this challenge, numerous frameworks and approaches have emerged. Soheil *et al.* proposed Network Deconvolution (ND) [1] for recognizing direct relationships between variables. They hold the idea that the noise is generated by a diffusion process which can be formulated as the power sum of direct relationships and develop an algorithm to eliminate the indirect effect. Silencer [2] model was also proposed by Baruch *et al.* to remove the indirect noise, which treats the perturbations of observed correlations as the cumulative result of local perturbations and removes the indirect effects by establishing the relationship between global and local perturbations. Babak *et al.* [4] pointed out that both approaches are related to the method of partial correlation and are essentially the scaled versions of the inverse correlation matrix. And the two models have been severely criticized by many scholars for their poor mathematics [5]. Nevertheless, the idea proposed by Soheil *et al.* [1] is very inspiring and similar models were used to understand the impact of network topology on human cognition, describing how humans learn and represent transition networks [6], [7]. Besides, Network Enhancement (NE) [3] was developed to improve the SNR of undirected, weighted networks by enhancing the signal intensity through a nonlinear operator which increases spectral eigengap. However, although plentiful denoising methods perform well on different datasets, most of them lack a comprehensive understanding of noise generation and are mostly based on heuristics thoughts while no rigorous mathematical interpretations are given to reveal the essential reason for why those algorithms work.

In this paper, we have proposed a novel network denoising framework named Network Refinement (NR) that takes a noisy network as input and outputs a network on the same vertex set with adjusted edge weights (enhanced signal or weakened noise) whose SNR is improved (Fig.1a, b). NR comprises two methods NR-F and NR-B based on whether it is the forward or backward process of the random walk diffusion model we have proposed (Fig.1c). Although NR-F and NR-B are built on different assumptions and denoising perspectives, they are unified into one denoising framework by a diffusion model, and we need to choose the appropriate denoising methods for the noisy networks according to the properties of networks themselves. Specifically, NR-F is built based on the intuition that edges between nodes connected by more paths and higher weights are more likely to be true edges (signal) for the noisy networks that have suffered from stochastic noise, while the intuition of NR-B is that the indirect edges (noise) are the cumulative transitive effect of direct edges for the noisy networks that have suffered from indirect effects. NR-F assumes that the self-organizing signals within networks can be strengthened by a diffusion process, while NR-B assumes that the observed network is derived from the diffusion of a true network that contains only direct effects. In other words, NR-F cares more about the properties of the signal while NR-B cares more about the reason for noise generation.

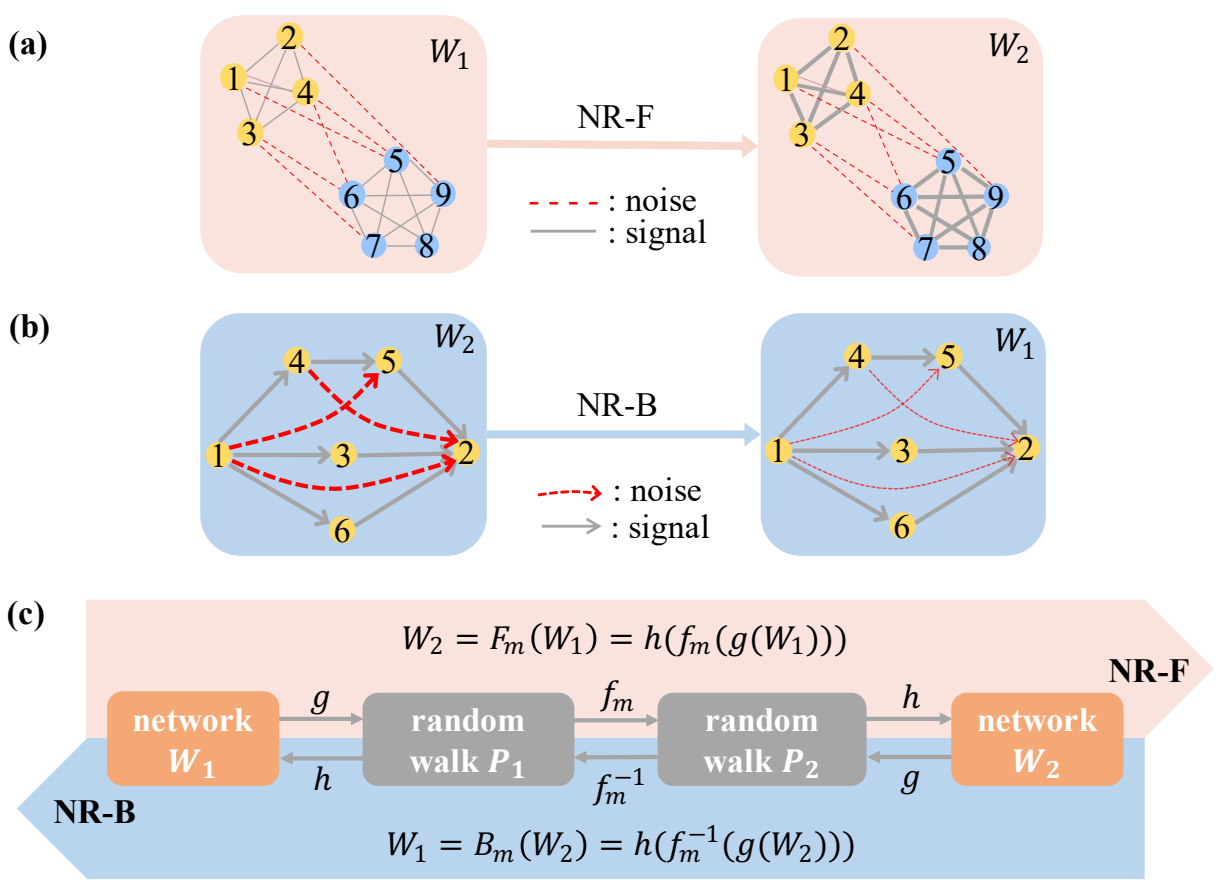

Fig. 1. Taking a noisy network as input, NR outputs a denoised network on the same vertex set with a new set of edge weights whose SNR is improved. Specifically, **(a)** when we apply NR-F on the noisy network, the denoised network enhances signal intensity; **(b)** when we apply NR-B on the noisy network, the denoised network reduces noise intensity. **(c)** The constituents of the composite graph operator $F_m$ of NR-F and the composite graph operator $B_m$ of NR-B, where the operator $f_m$ and $f_m^{-1}$ are the forward and backward process of the diffusion model defined by random walk on the graph, the operator $g$ and $h$ help us to implement the transformation between the graph and the random walk on the graph. When we apply operator $F_m$, we take $W_1$ as input and get the denoised network $W_2$; when we apply operator $B_m$, we take $W_2$ as input and get the denoised network $W_1$.

The main contributions of this work can be summarized as follows:

(1) First, we are the first systematic network denoising framework that attempts to denoise networks from different perspectives and integrates them using one diffusion framework defined by random walk on the graph.
(2) Second, our diffusion model is built on the random walk on the graph rather than on the graph directly, which avoids meaningless multiplication of the edge weights. We also discuss the connection and difference between the two diffusion models.
(3) Third, we give a one-to-one correspondence (ignoring constant multipliers) between the graph and the random walk on the graph, allowing us to map the diffusion process defined on the random walk to that defined on the graph.
(4) Fourth, we show that NR can significantly refine the quality of many networks by several applications on simulated networks and typical real-world biological and social networks: improving the accuracy of community detection; enhancing the quality of Hi-C contact maps from the human genome; distinguishing strong and weak relationships of social networks; refining the inference of gene regulatory networks. And the performance of NR is superior to other state-of-the-art network denoising methods. That confirms the importance of denoising network before doing network-based downstream analysis.

The paper is outlined as follows: Section II provides a detailed introduction of our network denoising framework, and the proofs of all the theorems in this section are given in Appendix A. Section III shows the results of applying NR to two simulated networks and four typical real-world networks. Some supplementary results of this section are given in Appendix B. Section IV gives a conclusion of our work.

## II. MODELS

Given a weighted undirected graph $G = (V, E, W)$ with vertex set $V$, edge set $E$ and edge weight matrix $W$, we can define a finite Markov chain with state-space $V$ and transition matrix $P = D^{-1}W$, where $D$ is the degree matrix of $G$, and we call this Markov chain a random walk on the graph $G$. In fact, there is not much difference between the theory of random walk on graphs and finite Markov chains if we treat the variable index $i$ and $j$ of Markov chain as the vertex index of the graph.

To formulate our NR framework, we will define six operators: $f_m$, $f_m^{-1}$, $g$, $h$, $F_m$ and $B_m$. The operators $f_m$ and $f_m^{-1}$ are related to the diffusion process of random walk on the graph, while the operators $g$ and $h$ help us map this diffusion to the underlying graph. The graph operator $F_m$ of NR-F is defined based on the operator $f_m$, and the graph operator $B_m$ of NR-B is defined based on the operator $f_m^{-1}$.

### A. *The diffusion operator $f_m$*

Denote $\boldsymbol{\mathcal{P}}$ as the set of transition matrices and $\boldsymbol{\mathcal{W}}$ as the set of (weighted) adjacency matrices, where $P \in \boldsymbol{\mathcal{P}}$ gives the representation of a random walk on a graph, and $W \in \boldsymbol{\mathcal{W}}$ gives the representation of a graph. The operator $f_m$ transforms a transition matrix $P$ to another by adding the probability of all paths of different length $k$ joining two nodes, with a smaller weight coefficient $1/m^k$ for a longer path of length $k$:

$$f_m(P) = \frac{\sum_k (P^k/m^k)}{\sum_k (1/m^k)} = (m-1)P(mI-P)^{-1} \tag{1}$$

where $\sum_k (1/m^k)$ is a normalization factor and $m > 1$.

It's easy to check that the spectral radius of $P/m$ satisfies $\rho(P/m) < 1$ when $m > 1$: the sum of each row of $P/m$ is $1/m < 1$, and for any non-negative matrix $A$, we have:

$$\min_{1\le i\le n} \sum_{j=1}^{n} a_{ij} \le \rho(A) \le \max_{1\le i\le n} \sum_{j=1}^{n} a_{ij}$$

so $\rho(P/m) = 1/m < 1$, which promise the convergence of the infinite series: $f_m(P) = (m-1)P(mI-P)^{-1}$.

Now we show that the operator $f_m$ transforms a transition matrix $P$ to another, as stated in Theorem 1:

*Theorem 1.* $f_m(P) \in \boldsymbol{\mathcal{P}}$ for $P \in \boldsymbol{\mathcal{P}}$. That's to say:
(1) The sum of each row of the matrix $f_m(P)$ is 1.
(2) $f_m(P)$ is a non-negative matrix.

Moreover, $f_m(P)$ is a polynomial of $P$, so it keeps many properties of $P$ unchanged. Formally, we have Theorem 2:

*Theorem 2.* Let $P$ be a transition matrix and $\{X_k\}$ be the random walk defined by $P$. The operator $f_m$ keeps the following properties of $P$ and $\{X_k\}$ unchanged:
(1) The reversibility of random walk $\{X_k\}$, thus the undirected property of the underlying graph.
(2) The stationary distribution of random walk $\{X_k\}$.

(3) Let $G$ be the graph where $\{X_k\}$ be treated as a random walk, then the degree distribution of $G$ keeps unchanged under operator $f_m$.

Besides, we can get a different form of $f_m(P)$ by simple calculations:

$$f_m(P) = a_1P + (a_2P^2 + a_3P^3 \dots) \triangleq P_o + P_d$$

where $a_k = (m-1)/m^k$. The first term $P_o = a_1P$ implies the impact of the original input transition matrix and the second term $P_d$ implies the impact given by a diffusion process on $P$. Thus, the proportion of the input transition matrix $P$ in the output transition matrix $f_m(P)$ is $a_1 = (m-1)/m$, which is a monotone increasing function when $m$ is greater than 1. It signifies that the parameter $m$ controls the modification extent to the input matrix $P$. In concrete words, we adjust the input transition matrix more gently with a bigger $m$, more violently with a smaller $m(m > 1)$. We set $m = 2$ as a default choice in the following numerical experiments, we will analyze the effect of parameter selection on the results of each numerical experiment.

In fact, the operator $f_m$ has described a process of signal enhancement by accumulating the transition probabilities of all lengths of paths between the two vertexes, which can also be regarded as a process of signal diffusion. Besides, the probability multiplication is more reasonable than edge weight multiplication in ND algorithm [1] for the representation of accumulative effect, because the probability multiplication indicates the simultaneity of events while edge weight multiplication has no practical meanings. Moreover, diffusion on the random walk can describe the graph evolution process for its mathematical explanation: If $P$ is the one-step transition matrix, then $P^k$ is the $k$-step transition matrix with $(P^k)_{ij}$ indicating the probability of $i$ connecting $j$ through a walk of length $k$.

### B. The inverse operator $f_m^{-1}$

Now we focus on the inverse process of the diffusion process defined by $f_m$. Considering that the diffusion will cause indirect effects, we can assume that the random walk on the noisy network is generated from the diffusion on a potential network, then we might make use of the inverse process of $f_m$ to eliminate the noise. That's to say, we can treat

$$P_{obs} = a_1P_{dir} + \left(a_2{P_{dir}}^2 + a_3{P_{dir}}^3 \dots\right)$$

where $P_{obs}$ represents the random walk on the observed noisy network, and $P_{dir}$ represents the random walk on the underlying true network. Then the inverse operator of $f_m$ can recover $P_{dir}$ from $P_{obs}$ by removing the indirect effect of all lengths of paths between two nodes.

In details, the inverse operator $f_m^{-1}$ of $f_m$ is defined as follows:

$$f_m^{-1}(P) = m\big((m-1)I + P\big)^{-1}P \tag{2}$$

It's easy to verify that Theorem 3 holds:

*Theorem 3.* The operators $f_m$ defined in (1) and $f_m^{-1}$ defined in (2) are inverse operators to each other.

Notice that the operator $f_m^{-1}$ has described a reverse process compared to $f_m$, which removes the effect of all indirect paths joining two nodes, but shares the same properties as the operator $f_m$ stated in Theorem 2:

*Theorem 4.* Let $P$ be a transition matrix and $\{X_k\}$ be the random walk defined by $P$. The operator $f_m^{-1}$ keeps the following properties of $P$ *and* $\{X_k\}$ unchanged:
(1) The reversibility of random walk $\{X_k\}$, thus the undirected property of the underlying graph.
(2) The stationary distribution of random walk $\{X_k\}$.
(3) Let $G$ be the graph where $\{X_k\}$ be treated as a random walk, then the degree distribution of $G$ keeps unchanged under operator $f_m^{-1}$.

Besides, we need to promise that the operator $f_m^{-1}$ also change a transition matrix to another, that's to say: $f_m^{-1}(P) \in \boldsymbol{\mathcal{P}}$ for $P \in \boldsymbol{\mathcal{P}}$. Firstly, we have:

*Theorem 5.* The sum of each row of the matrix $f_m^{-1}(P)$ is 1 for $P \in \boldsymbol{\mathcal{P}}$.

However, the nonnegativity of $f_m^{-1}(P)$is not always guaranteed, that's because the basic assumption of using the operator $f_m^{-1}$ to remove all indirect effects is that the random walk on the noisy input network is (approximately) generated by the diffusion process defined by the operator $f_m$, when the truth is far away from this assumption, the inaccuracy could lead to abnormal results. Especially, it's unimaginable to remove the indirect effects of a sparse network, and the negative numbers occur when there is no edge between two nodes or the weight of two nodes is too small to burden the sum of all paths of different lengths joining them. To solve this ill-conditioned problem which prevents us from making sure that $f_m^{-1}(P) \in \boldsymbol{\mathcal{P}}$, we will give a revised method in the following subsection $D$.

### C. Two auxiliary operators $g$ and $h$

Next, we will define two auxiliary operators $g$ and $h$ which realize the transformation between the graph and the random walk on the graph. These two operators help us mapping the diffusion process of the random walk defined by the operator $f_m$ and $f_m^{-1}$ to the diffusion process on the graph.

Denote $D$ as the diagonal degree matrix of $W$, then $g(W)$ defines a random walk on the graph whose (weighted) adjacency matrix is $W$:

$$g:\ \boldsymbol{\mathcal{W}} \to \boldsymbol{\mathcal{P}}$$
$$g(W) = D^{-1}W \tag{3}$$

The operator $h$ has the opposite effect of $g$, which recovers the underlying graph of the random walk defined by the transition matrix $P$:

$$h:\ \boldsymbol{\mathcal{P}} \to \boldsymbol{\mathcal{W}}$$
$$h(P) = \alpha \cdot \mathrm{diag}(\pi(P))P \tag{4}$$

where $\pi(P) = (\pi_1, \dots, \pi_n)$ is the stationary distribution of transition matrix $P$ such that $\pi P = \pi$. When $P$ defines an irreducible and aperiodic random walk, $\pi(P)$ exists and can be guaranteed to be unique (that is generally true in practical applications). $\mathrm{diag}(x)$ means the diagonal matrix whose diagonal element is the vector $x$, and $\alpha$ is a constant which controls the sum of weight matrix $h(P)$. The operator $h$ multiplies the transition probability $P_{ij}$ by the stationary distribution of node $i$ which reflects the degree information of the graph.

The operators $h$ and $g$ have established a one-to-one correspondence between the graph and the random walk on the graph (ignoring constant multipliers), as stated in Theorem 6:

*Theorem 6.* The operators $g$ defined in (3) and $h$ defined in (4) are inverse operators to each other.

### D. NR denoising model

Formally, the NR denoising framework is defined based on the operators we have proposed above, which changes one graph into another by adjusting the edge weights. The pipeline of NR is shown in Fig.1 c.

For NR-F, to map the diffusion process of random walk on the graph defined by $f_m$ onto the diffusion process on the graph, we wrap the operator $f_m$ in operators $g$ and $h$ to get the composite operator $F_m$:

$$F_m\colon \boldsymbol{\mathcal{W}} \to \boldsymbol{\mathcal{W}}$$
$$F_m(W) = h(f_m(g(W))) \tag{5}$$

where operator $g$ and $h$ realize the conversion between the graph and random walk on the graph, and $f_m$ realize the diffusion of random walk on the graph. The operator $F_m$ describes the process of signal diffusion on the graph, which avoids the meaningless multiplication of edge weights directly. In fact, the NR-F model defined by the operator $F_m$ has the following relationship with the ND algorithm [1] that defines the diffusion model on the graph directly:

*Theorem 7.* For any input matrix $W$, the $m$-th order edge weight matrix of $F_m(W)$ defined by the $m$-th step transition matrix of $g(W)$ is (ignoring constant multiplies):

$$\sum_{k_1,k_2,\dots k_{m-1}} \left( \frac{W_{ik_1} W_{k_1k_2} \cdots W_{k_{m-1}j}}{\sum_l W_{k_1l} * \sum_l W_{k_2l} \cdots * \sum_l W_{k_{m-1}l}} \right)$$

which is a degree normalized version of the $m$-th order edge weight matrix of ND algorithm:

$$\sum_{k_1,k_2,\dots k_{m-1}} \left( W_{ik_1} W_{k_1k_2} \cdots W_{k_{m-1}j} \right)$$

Thus, our NR-F model actually normalizes the product of the edge weights by the degree of the nodes in the path, which makes the path with higher intermediate node degrees less important because their information is dispersed through more adjacent edges. Although similar formulas have been used before [8] in a local way for unweighted graphs, which is a special case of our model, we provide a different angle for understanding them and explain why it works more essentially.

Similarly, the inverse process defined by $f_m^{-1}$ can also be mapped onto the graph. We define this composite graph operator as follows:

$$B_m\colon \boldsymbol{\mathcal{W}} \to \boldsymbol{\mathcal{W}}$$
$$B_m(W) = h(f_m^{-1}(g(W))) \tag{6}$$

The operator $B_m$ aims at removing the combined indirect effects of all lengths of paths connecting two nodes. Assuming that the noisy network is approximately generated by the diffusion process defined by the operator $F_m$, then hopefully we could utilize the operator $B_m$ to remove all the indirect effects.

We have mentioned before that the matrix $f_m^{-1}(g(W))$ may not be nonnegative, which prevents it from being a standard transition matrix. To solve this ill-conditioned problem, we preprocess the input network by modifying it to $\widetilde{W} = W + \varepsilon_1 J + \varepsilon_2 I$ before applying NR-B, where $J$ is the matrix whose entries are all 1, $I$ is the identity matrix whose diagonal entries are 1, $\varepsilon_1$ and $\varepsilon_2$ are the parameters that control the level of the modification to the input network.

Namely, when the input network is unweighted, we change the input adjacency matrix into a weighted adjacency matrix by strengthening the weights of all edges:

$$\tilde{A}_{ij} = \begin{cases} \varepsilon_1 + \varepsilon_2 & i = j \\ \varepsilon_1 + 1 & i \neq j,\ i \sim j \\ \varepsilon_1 & else \end{cases} \tag{7}$$

This preprocessing step can be understood as a recovery mechanism if we assume that the input network is obtained by thresholding a network and removing the diagonal entries. This harmless modification does not destroy the structure of the original network and helps to make $f_m^{-1}(g(W))$ non-negative if we take large enough values for the parameters $\varepsilon_1$ and $\varepsilon_2$. In order not to affect the properties of the input network, we take $\varepsilon_1 = \varepsilon_2 = 1$ for the unweighted graph and the minimum non-zero value for the weighted graph.

However, when the input network is weighted, preprocessing like $\widetilde{W} = W + \varepsilon_1 J + \varepsilon_2 I$ is usually not powerful enough to ensure the non-negativity of $f_m^{-1}(g(\widetilde{W}))$ due to the complexity of the distribution of the input network edge weights, thus we take a reasonable postprocessing step to the matrix $f_m^{-1}(g(\widetilde{W}))$ to make it non-negative. Specifically, we minus the smallest negative value from each row with negative numbers from $f_m^{-1}(g(\widetilde{W}))$ to eliminate negative entries. Denote $f_m^{-1}(g(\widetilde{W}))$ as $P_B$, then the postprocessing step can be written as:

$$\widetilde{P_B} = P_B - (\beta_1 \mathbf{1}, \dots \beta_n \mathbf{1})^T \tag{8}$$

where $\mathbf{1}$ is the column vector whose elements are all 1, $\beta_i = 0$ if the $i$-th row of $P_B$ has no negative elements, $\beta_i = \min_i\{(P_B)_{ij}\}$ if the $i$-th row of $P_B$ has at least one negative element.

The detailed revised pipeline of NR framework has been given in Fig. S1 of Appendix B.

## III. Applications

To test the effectiveness of NR framework for network denoising, we apply it to two types of simulated networks and four typical real-world biological and social networks.

The simulated networks are generated according to the denoising principle of NR-F and NR-B respectively, then we use the corresponding denoising method to improve their SNR. In each experiment, we evaluate the performance of NR and two comparative methods: ND [1] and NE [3]. Since NE aims at enhancing signals and ND is developed for removing indirect noise, we mainly compare NR-F with NE and compare NR-B with ND. Notice that NR has a parameter $m$ that controls the adjustment intensity of the edge weights. We set $m = 2$ as a default choice, and we will show the sensitivity of $m$ in different applications.

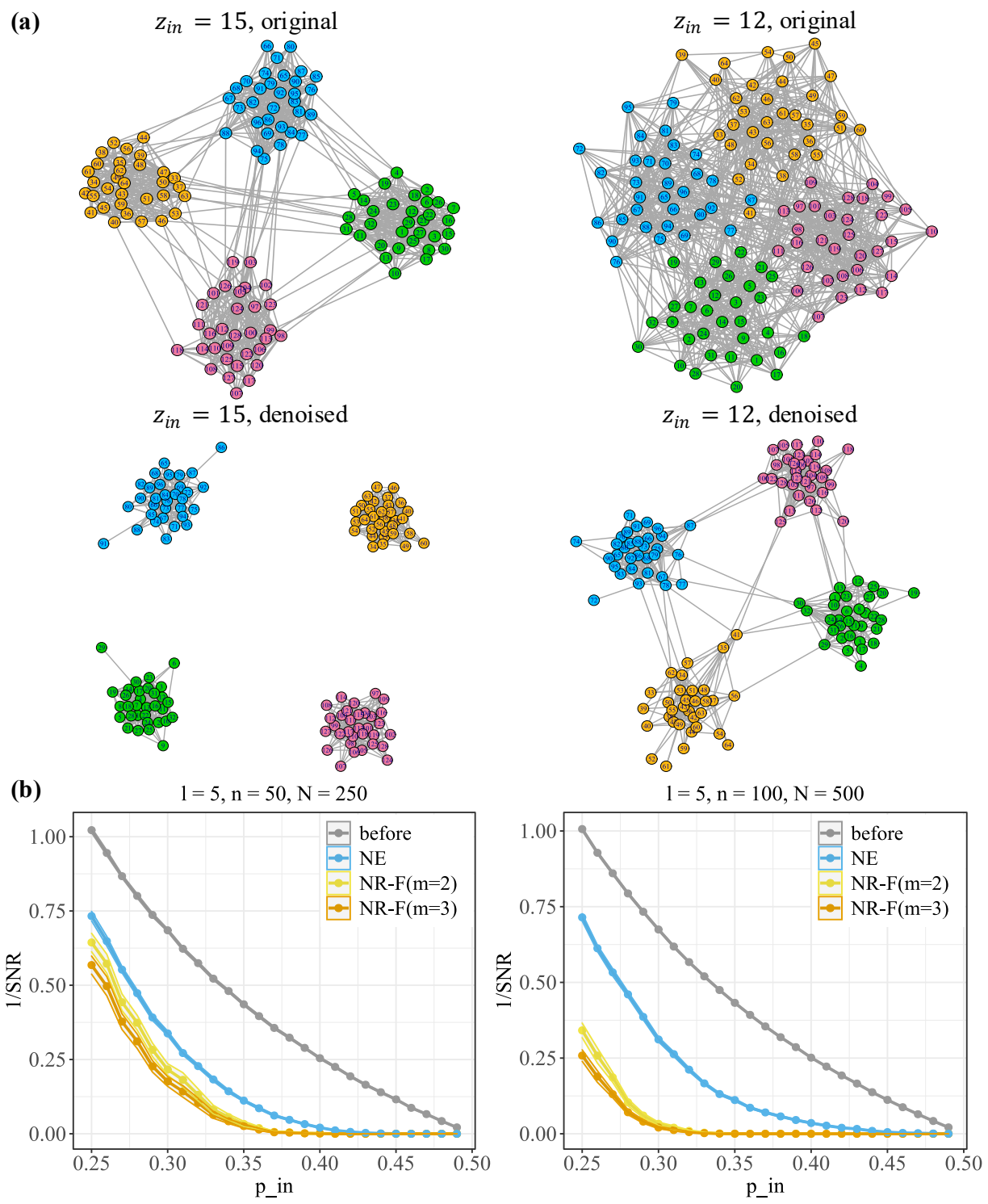

Fig. 2. **(a)** The structure of the benchmark networks given by Girvan and Newman before and after applying NR-F. We show the performance of two special cases when $z_{in} = 15$ and $z_{in} = 12$. Notice that we choose an appropriate threshold to the denoised network such that the unweighted network has no isolated vertices. **(b)** Evaluation of denoising performance by comparing the $1/\text{SNR}$ before and after applying NE and NR-F. We control the noise intensity by taking different values of $p_{in} \in [0.25,0.5]$ and test in two network configurations: N=250 and N=500. Same as before, we choose the threshold such that the unweighted network has no isolated vertices to compute the $1/\text{SNR}$.

### *A. NR increases SNR of two types of simulated noisy networks*

To test the effectiveness of NR-F, we simulate networks with well-defined signals so that we can improve their SNR by enhancing their signal intensity. We choose the standard benchmark networks for community detection which are generated with the planted $l$-partition model [9]. The model partitions a network with $N = nl$ vertices into $l$ groups with $n$ vertices each. Vertices of the same group are linked with probability $p_{in}$, whereas vertices of different groups are linked with probability $p_{out}$. The expected internal and external degree of a vertex can be calculated as $z_{in} = p_{in}n$ and $z_{out} = p_{out}n(l-1)$ [9]. The community structure of a network is well defined when $z_{in} > z_{out}$. The inter-group edges are regarded as noise while the other edges are considered as signals.

Girvan and Newman [11] considered a special case of the planted $l$-partition model where they let $n = 32$, $l = 4$, and $p_{in} + (l-1)p_{out} = 1/2$. We generate such networks with parameters $z_{in} = 15$ and $z_{in} = 12$, then we apply NR-F to them to get the denoised networks (Fig. 2a). Because the edge weights inside groups are enhanced by NR-F, we can get a new network exhibiting clearer community property based on a reasonable threshold (we choose the threshold such that the unweighted network has no isolated vertices for better presentation). The result shown in Fig. 2a demonstrates the effectiveness of NR-F to improve the SNR of networks.

We further comprehensively compare the performance of NR-F and NE on the benchmark networks at different noise levels in two different network configurations: $N = 250$ and $N = 500$. To make sure the network's community structure is well defined, we need to ensure that $z_{in}/z_{out} \geq 1$, which implies that $p_{out} \in [0,1/4(l-1)]$ and $p_{in} \in [0.25,0.5]$. Different values of $p_{in}$ generate different levels of noise, where smaller $p_{in}$ gives a stronger noise and vice versa. When $p_{in} = 0.25$, the network has lost the community properties; when $p_{in} = 0.5$, the network has $l$ connected components and can be directly regarded as $l$ communities. The denoising performance is evaluated by comparing 1/SNR before and after applying NR-F and NE (since the SNR is infinite when the number of noise edges is 0, we show 1/SNR instead of SNR). The results for each network configuration are the average of 50 experiments. Fig. 2b shows that $1/\text{SNR}$ decreases significantly after applying NR-F for $m = 2,3$, and its denoising effect is better than NE, especially when the network is relatively large (denoising task is more difficult). We also test the performance of NR-F under different levels of denoising difficulties ($n = 50, l = 4,5,6,7$ and $n = 100, l = 4,5,6,7$), the similar denoising results are shown in Fig. S2 of Appendix B.

To evaluate the ability of NR-F to identify the signal edges, we also calculate the area under Precision-Recall curves (AUPR) and the area under Receiver Operating Characteristic curves (AUROC) scores, where we treat intra-group edges as positive labels and the remaining edges (including inter-group edges and all absent edges) as negative labels. The results are shown in Fig. S3 of Appendix B. We can see that the scores of NR-F are consistently higher than NE under different evaluation metrics and network configurations.

To test the effectiveness of NR-B, we simulate the noisy networks by adding indirect edges to cycles, Erdős–Rényi random graphs and BA graphs [12], where the indirect edges are defined as the edge between nodes that are path-connected in the original network and the two nodes connected by a shorter path are more likely to add an indirect noisy edge. Then we apply NR-B to the noisy network to obtain the denoised weighted network and compare their 1/SNR.

We first generate the circle graph with 20 nodes as the true network and add indirect noisy edges to it to get the noisy network, then we apply NR-B and ND to the noisy network and get the denoised network, as shown in Fig. 4a. This noisy network contains 20 noisy edges, while NR-B denoised network has only 4 noisy edges and ND denoised network has 10 noisy edges (we keep as many edges as the original network). By comparing the number of noisy edges marked in red of the original network and the denoised network, we demonstrate the effectiveness of NR-B in noise elimination.

We further comprehensively compare the performance of NR-B and ND on Erdős–Rényi (ER) random graphs of node size $N = 50$, edge-formation probability $p = 0.3$ and Barabási–Albert (BA) graphs of node size $N = 50$, $m = 3$ (the number of edges to add in each time step). Then we add indirect noisy edges of different intensities to those graphs, where the proportion of noisy edges ranges from 0.1 to 0.5. We apply NR-B and ND to the noisy graphs to denoise them and compare 1/SNR scores of noisy and denoised networks. For each proportion of noisy edges, we show the average score of 100 repeated tests. Same as the results of NR-F, 1/SNR decreases significantly after applying NR-B for $m = 2$ and $m = 3$, and its denoising performance outperforms that of ND (Fig. 4b).

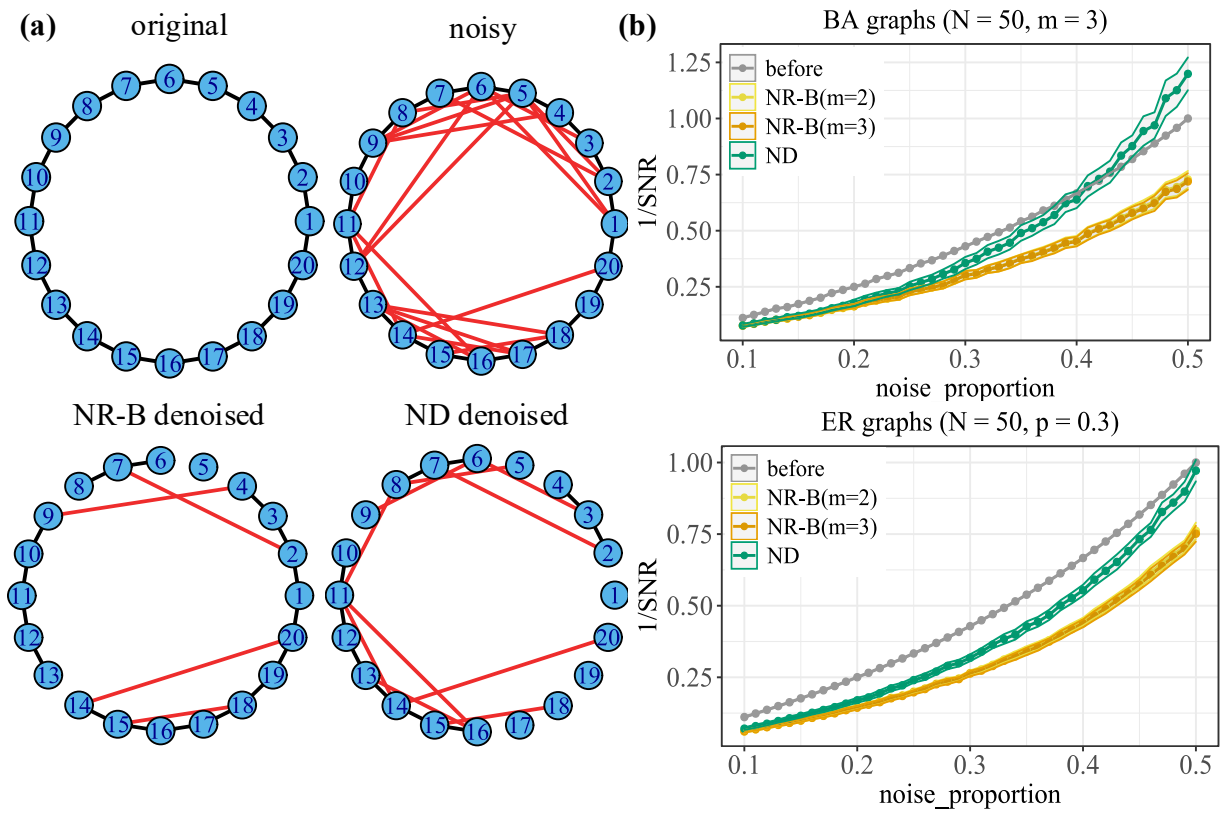

Fig. 4. **(a)** The original network is a circle graph with 20 nodes and the noisy graph is obtained by adding indirect edges to the true network with the noisy edges marked by red color. We apply NR-B and ND to the noisy network to improve its SNR (we take the threshold such that the denoised network has the same number of edges as the original network for better presentation). The number of noisy edges of NR-B denoised network and ND denoised network is 4, 10 respectively. **(b)** Evaluation of denoising performance by comparing the $1/\mathrm{SNR}$ before and after applying ND and NR-B. We control the noise proportion (number of noise edges/total number of edges) ranging from 0.1 to 0.5, and test in noisy BA graphs and ER graphs. Same as before, we choose the threshold such that the unweighted network has the same number of edges as the original network to compute the $1/\mathrm{SNR}$. The yellow line and the orange line almost overlap because NR-B is insensitive to the value of the parameter $m$.

We also compute the AUPR and AUROC scores to show the recovery accuracy of NR-B (Fig. S4 of Appendix B). We can see that the scores of NR-B are consistently higher than ND under AUPR and AUROC evaluations in different noisy graphs, and have a smaller variance which shows the robustness of NR-B.

### *B. NR-F improves the accuracy of community detection*

Many networks exhibit community structure with certain nodes showing strong self-organization properties [9], [10], [13]-[15]. Community detection in networks, also called graph or network clustering may offer insight into how networks are organized. It has been shown the importance of denoising networks before feeding them into clustering algorithms [16]. We can treat the edges within the community as signals and assume that the network suffers from stochastic noise that makes the network structure unclear, then we demonstrate that NR-F can be used to enhance their signal intensity and reinforce their community structure. The network we have tested is the well-known Zachary's karate club network [17], which consists of 34 vertices representing the members of the club, and the edge between vertices indicates the relationship between members. A conflict splits the club into two groups gathering around vertex 34 (the president) and 1 (the instructor), as shown in Fig. S5 of Appendix B.

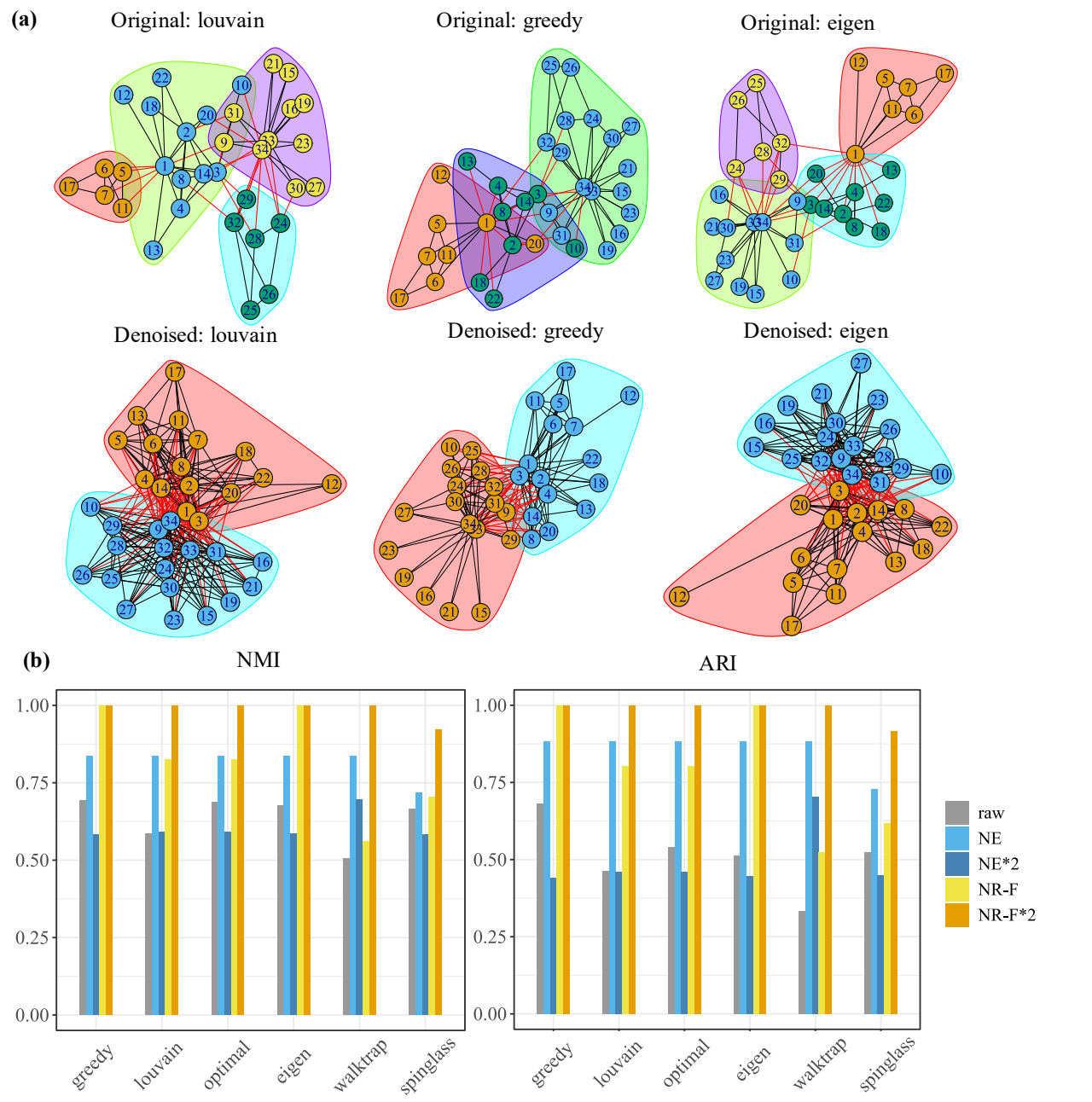

Fig. 3. **(a)** The community detection results before (top) and after (bottom) taking NR-F as a preprocessing step. We use different colors to mark different groups. Here we only show the clustering performance of three methods: Louvain algorithm [19], fast-greedy algorithm [14], and leading-eigen algorithm [13]. For display purposes, we set an appropriate threshold to change the weighted network into an unweighted network. **(b)** Improvement of the NMI and ARI evaluations on Zachary's network after applying NR-F and NE once and twice for six community detection methods (all the community detection methods can be found in R packages *igraph*).

There are numerous methods have been proposed to detect communities of networks. We evaluate the clustering performance of six community detection methods (fast-greedy [14], optimal [18], Louvain [19], leading-eigen [13], Walktrap [15], spinglass [20]) on Zachary's network before and after applying NR-F and NE as a preprocessing step. We use Rand Index (RI) [21], Adjusted Rand Index (ARI) [22], and Normalized Mutual Information (NMI) [23] to evaluate the performance of those graph clustering methods. The results shown in Fig. 4a graphically illustrate the clustering results of three community detection methods (fast-greedy [14], Louvain [19], leading-eigen [13]) before and after applying NR-F, and similar results can be obtained with the other three methods (Fig. S6 of Appendix B). Besides, Fig. 4b shows the improvement of ARI and NMI evaluations after applying NR-F and NE once and twice, and RI evaluation shows similar results (Fig. S7 of Appendix B). When we apply the community detection algorithms on the original Zachary's network, the predicted clustering labels of various methods are far from the real labels. When we adjusted the edge weights of Zachary's network by NR-F, the clustering performance of various methods has been greatly improved. Besides, we can see from Fig. 4b that applying NR-F repeatedly can better improve the clustering performance. The classification results of most community detection methods are exactly right on the network that has been applied NR-F twice while applying NE twice gets worse performance.

### *C. NR-F enhances the quality of Hi-C contact maps from the human genome*

The technique of high-throughput chromosome conformation capture (Hi-C) reveals the spatial properties of the human genome [24], [25]. Hi-C data are generally represented as a contact matrix $M_{n\times n}$, where the genome is partitioned into $n$ equally sized bins and $M_{ij}$ is the number of normalized counts of reads mapping to genomic regions $i$ and $j$ [26]. The bin size decides the resolution of Hi-C map, and the Hi-C map with a higher resolution gives more information and finer chromosome structure, yet higher sequencing cost. Thus, most available Hi-C maps have relatively low resolution [27], which prevents us from identifying some refined structures of 3D genome organization. Therefore, it's valuable to develop computational models for improving the quality of the Hi-C map. It has been demonstrated that the genome can be partitioned into higher-order A/B compartments, which are further partitioned into topologically associating domains (TADs) and smaller, nested sub-TADs [28], [29]. Based on the structural properties of the chromosomes, we can assume that the signal of Hi-C data is self-organizing, thus we can enhance its signal intensities by NR-F.

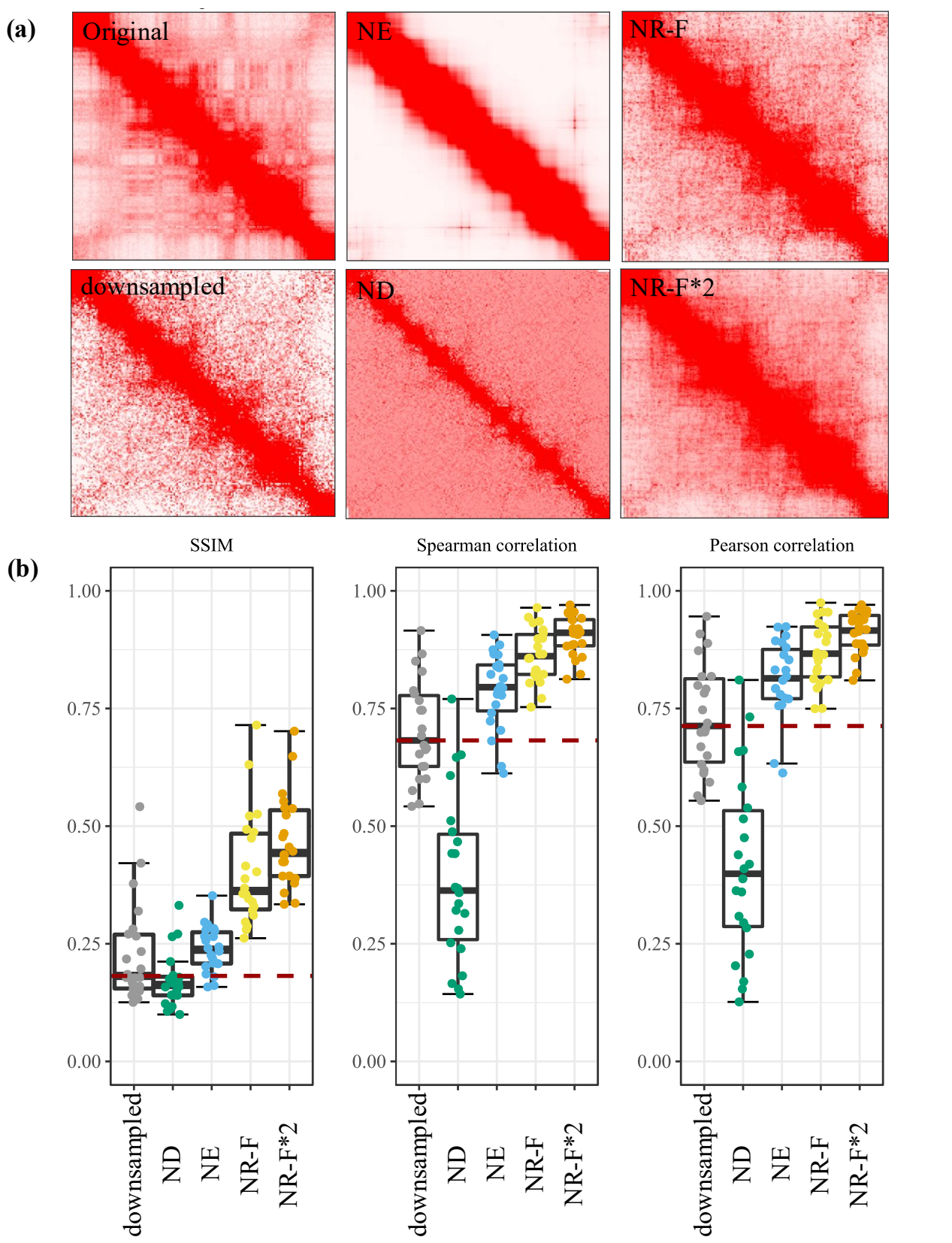

Fig. 5. **(a)** Enhanced Hi-C matrices and original Hi-C matrix in the 18.8Mb-36.2MB fragments of chromosome 13 of GM12878 (174×174 bins). Six panels are showing the fragments of different Hi-C contact matrices, which are marked by the title at the top of each panel. **(b)** The denoising performance of NR-F, NE, and ND under three evaluations (SSIM, Pearson correlation, and Spearman correlation). We apply three algorithms to denoise the downsampled matrices and compare them with the original contact matrix. Notice that the scores are all calculated by sliding sub-windows between matrices in the diagonal.

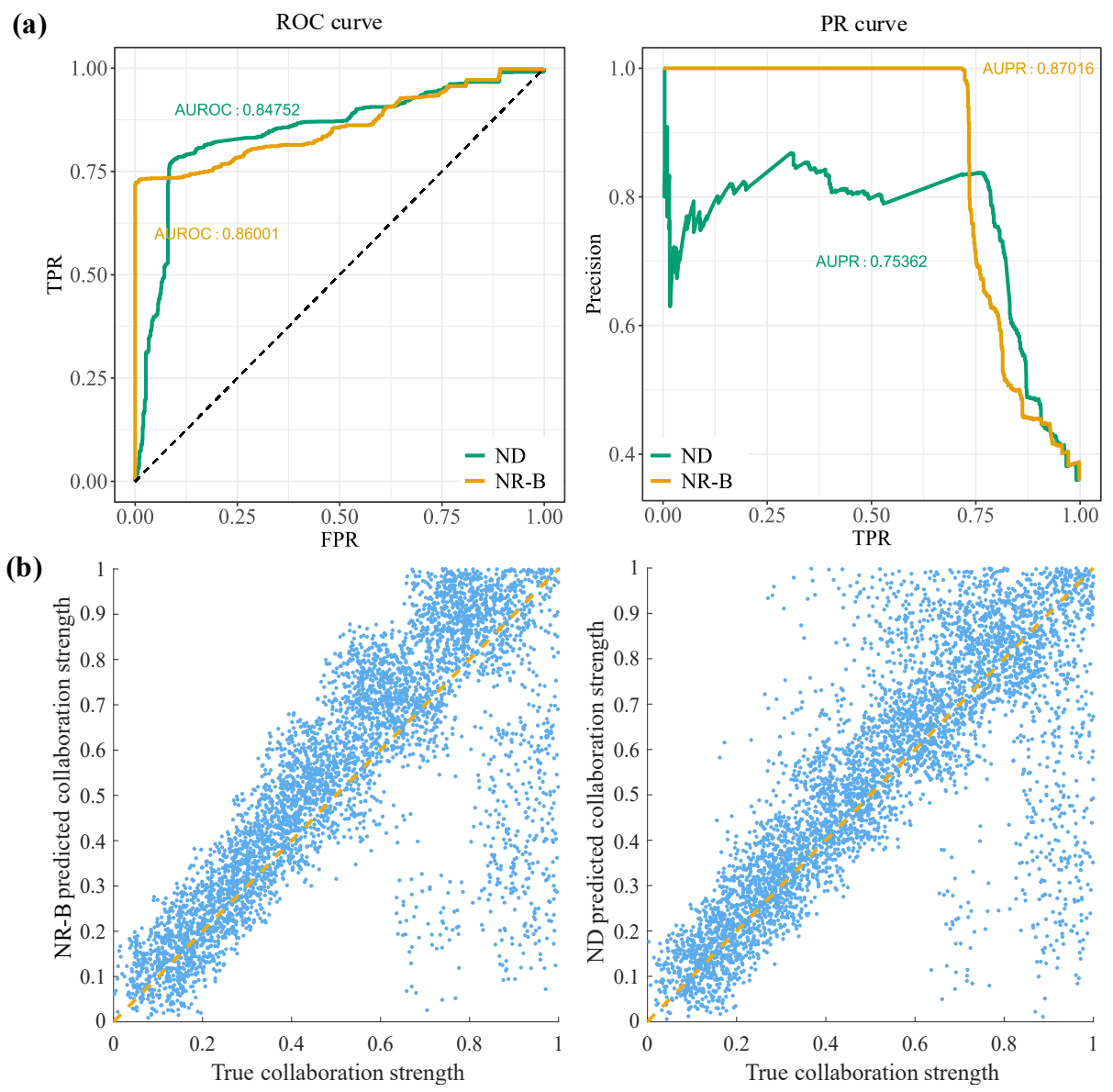

Fig. 6. **(a)** The ROC curve and PR curve of NR-B and ND predicted collaboration strengths with ND shown in green and NR-B shown in orange. **(b)** The correlation of Newman's collaboration strength and the predicted strength of ND and NR-F. For display purposes, the edge strength we have used is the transformation of the rank value of the edge other than the edge weight directly.

To facilitate the comparison of Hi-C matrices with different resolutions, contact matrices with lower sequencing depth were used to simulate the lower resolution matrices [26]. We will show that although using only $1/100$ of the original sequencing reads, NR-F can obtain contact matrices quite similar to the original ones.

To be specific, we first get the 100kb resolution Hi-C matrix from the Hi-C experiments of Rao *et al*. [25], which can be downloaded from GEO (https://www.ncbi.nlm.nih.gov/geo/) under accession number GSE63525, we choose the primary Hi-C data in GM12878 B-lymphoblastoid cells (HIC001-018) with reads mapping quality $> 30$ and normalize them with the KR-normalization scheme [26]. And we get the corresponding low-quality Hi-C matrix by randomly downsampling its sequencing depth to 1/100. This low-quality Hi-C matrix has lost many signals because of a shallower sequencing depth. We then apply NR-F, ND, and NE to denoise this low-quality matrix and compare it with the original one. Fig. 5a graphically shows the denoising effect on the $18.8\text{Mb} - 36.2\text{MB}$ fragments of chromosome 13. Notice that instead of denoising the entire matrix directly, we only denoise the overlapping sub-matrices in the diagonal, as shown in Fig. S9 of Appendix B. If we get different values in different sub-matrices, we take their average.

To quantitatively compute the improvement, we compare the denoising performance of NR-F, ND, and NE under the evaluations of Structural Similarity (SSIM), Spearman correlation, and Pearson correlation calculated by sliding sub-window between matrices in the diagonal (Fig. S10 of Appendix B), in which the SSIM is a standard metric to measure the similarity of two Hi-C images [30]. For each matrix, outliers are set to the allowed maximum by using the threshold of the 95-th percentile [26], and all elements are rescaled by min-max normalization to [0,1] for comparison (The cutoff is to prevent outliers from destroying the normalization).

As shown in Fig. 5b, NR-F greatly improves the SNR of the downsampled Hi-C map and has a better performance than other methods. And same as the previous experiments, applying NR-F repeatedly exhibits better performance. The same result of NR-F can be obtained when $m = 3$ and $m = 4$ (Fig. S8 of Appendix B).

| #Kept_edges | GENIE3 | GENIE3 + ND | GENIE3 + NR-B |
|---|---|---|---|
| 100 | 97 | 99 | 100 |
| 200 | 177 | 191 | 200 |
| 300 | 255 | 275 | 299 |
| 400 | 334 | 354 | 395 |
| 500 | 405 | 429 | 490 |
| 1000 | 711 | 750 | 869 |
| 1200 | 794 | 851 | 976 |
| 1500 | 901 | 942 | 1113 |

Table 1. Comparison of the number of true positive edges when different numbers of edges are retained on the NR-B and ND denoised network. For the original weighted network inferred by GENIE3 and the weighted networks obtained by applying NR-B and ND on this network, we keep their top 100, 200, 300, 400, 500, 1000, 1200, 1500 highest scoring edges and compare the number of correctly inferred edges. We can see that NR-B greatly improves the network inference accuracy and has a better performance than ND.

### *D. NR-B distinguishes strong collaborations of co-authorship networks*

Social networks describe the relationships between individuals with strong and weak relationships playing different key roles [31]. The co-authorship network is a typical social network that describes the collaborative relationships among different scientists [32], [33]. The unweighted co-authorship network is established by connecting the edge if the scientists have co-authored at least one paper and vice versa [1]. The weighted co-authorship network is defined by using additional information including the number of co-authored papers and the total number of authors in each publication. Our basic assumption is that the true weight of edges between two nodes should be lower than the observed ones if there are more common paths connecting two nodes, considering that a large part of the observed edge weights is due to indirect effects caused by common paths. We assume that the strong relationships connect active collaborators, while the weak relationships are driven by active collaborators. For example, if A and B collaborate frequently, A and C collaborate frequently, then B and C are likely to be co-authors by chance. Thus, we can apply NR-B to distinguish true strong and weak relationships of unweighted co-authorship networks. The datasets we have tested were created by M. Newman [13] which contains 5484 edges among 1589 authors. We follow the definition given by Newman to define the strong collaborations as the edges whose weight ≥ 0.5 (36% of edges), and the remaining edges are defined as weak collaborations.

We apply NR-B and ND to the unweighted co-authorship network and get a weighted output network. We treat the problem of distinguishing strong and weak collaborations as a binary classification problem where we take strong and weak relationships as two labels to classify. We calculate the AUROC and AUPR scores to evaluate the performance of NR-B and ND, the results are shown in Fig. 6a.

We also plot the correlation of Newman's collaboration strength and the predicted strength of NR-B and ND in Fig. 6b. For display purposes, the edge strength we have used is the transformation of the rank value of the edge other than the edge weight directly, and more details can be found in ND [1]. It's easy to find that ND suffers from severe false positive problems while NR-B does not (taking strong collaborations as positive labels and weak collaborations as negative labels), which is also verified in Fig. 6a.

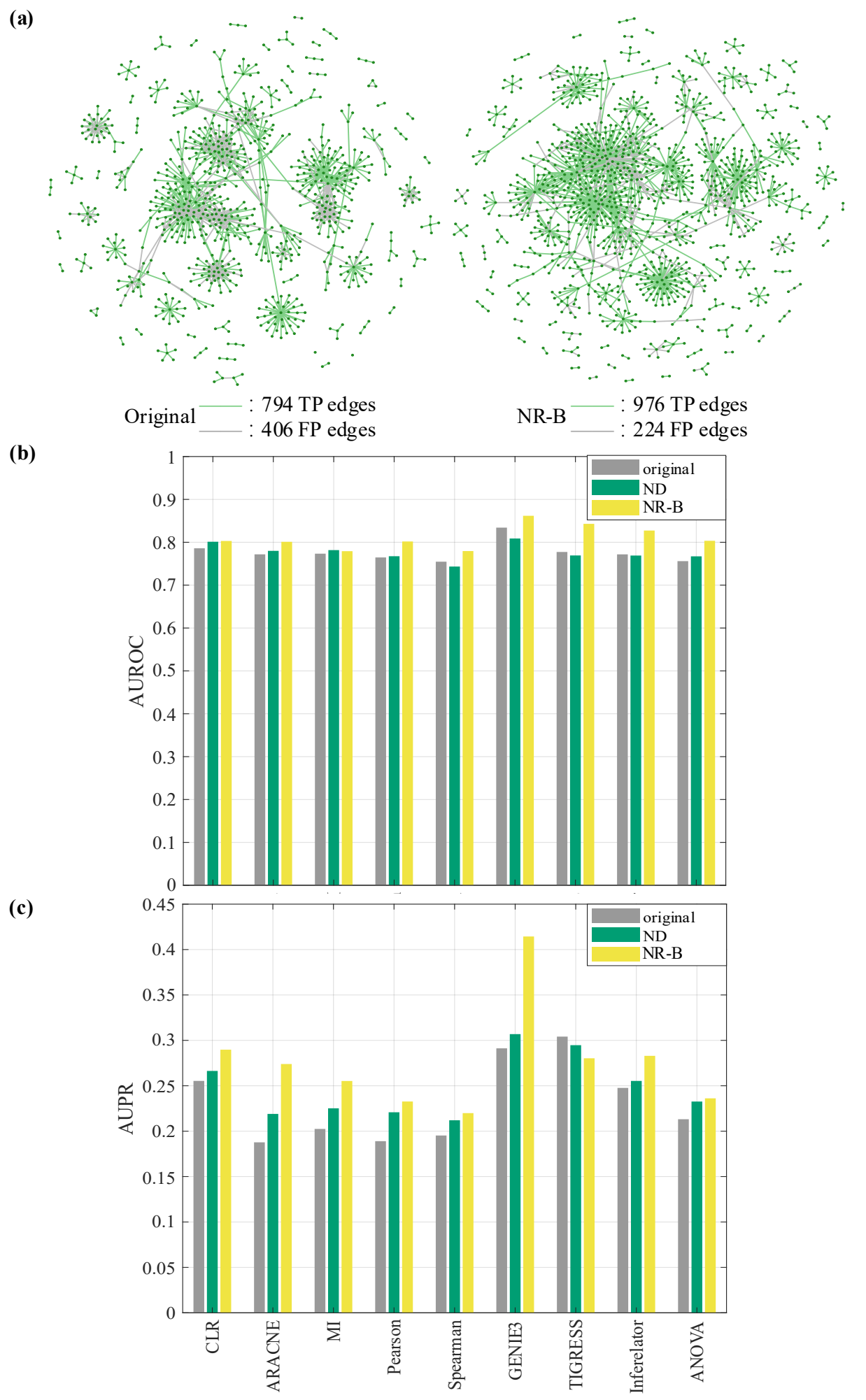

Fig. 7. **(a)** For the original weighted network inferred by GENIE3 and the weighted network obtained by applying NR-B on this network, we keep their top 1200 edges with the highest scores and compare the number of correctly inferred edges. We can see that 794 edges of the original network are correctly inferred while 974 edges of the NR-B denoised network are correctly inferred. **(b)** Improvement of the AUROC and **(c)** AUPR evaluations on the DREAM5 networks. As with ND, we have evaluated the noisy networks from nine GRN inference methods, in which CLR, ARACNE, MI, Pearson, Spearman are correlation-based methods, and GENIE3, TIGRESS, Inferelator, ANOVA are some top-performing inference methods from various felids. There are three colored bars for each method of each diagram: the grey one represents the performance of the method itself, the green one represents the performance after applying the ND algorithm to the noisy network inferred by the method and the yellow one represents the performance after applying NR-B.

### *E. NR-B refines the inference of gene regulatory networks*

Gene regulatory networks (GRN) are commonly used to describe the regulatory relationships between transcription factors and their target genes [34]-[36]. The boom in computational biology has made extrapolating GRN from data a hot topic [34]-[37], but when networks are constructed from data through statistical inference, it's likely to suffer from false-positive problems due to the transitive effects of correlation [1]. For example, if gene A regulates gene B and gene B regulates gene C, then there will be a strong correlation between the expression of genes A and C. Thus, the observed network can be seen as a diffusion of the true network, therefore we can use NR-B to denoise the inferred GRNs to improve their inference accuracy.

We test the GRNs constructed from the *in silico* gene expression datasets of the Dialogue on Reverse Engineering Assessment and Methods (DREAM) project [37], which is a comprehensive platform for evaluating the performance of various GRN inference methods. We demonstrate the effect of NR-B by comparing the inference performance of nine individual GRN inference methods before and after taking NR-B as a postprocessing step. These methods include correlation-based algorithms (CLR [38], ARACNE [39], MI, Pearson, Spearman), machine learning-based approaches (GENIE [40], TIGERSS [41]), and statistical approaches (Inferelator [42], ANOVA [43]).

For the network inferred by GENIE3 and the network obtained by applying NR-B on this network, we keep their top 1200 edges with the highest scores and compare the number of correctly inferred edges (Fig. 7a). We can see that 794 edges of the original network are correctly inferred while 974 edges of the NR-B denoised network are correctly inferred. The comparison of the true positive edges of ND and NR-B when 100-1500 edges are retained can be found in Table 1 and Fig. S11 of Appendix B.

Besides, by comparing the AUPR and AUROC scores with and without taking NR-B as a denoising step on the GRNs inferred from various methods, we show that NR-B can significantly promote the accuracy of nearly all network inference methods, and have a better performance than ND, as shown in Fig. 7b, c. The AUROC score has increased by 4.6% overall for NR-B while reduced to 99.9% for ND. Meanwhile, the AUPR score has increased by 17.9% overall for NR-B while increased by 7.0% for ND. We take $m = 2$ for NR-B in this experiment, while different choices of parameter $m$ for NR-B show the similar performance (Fig. S12 of Appendix B).

## IV. Conclusion

In this paper, we have proposed a novel network denoising framework called NR which consists of two methods named NR-F and NR-B, based on it is a forward or backward process of the diffusion model we have proposed. When the network has suffered from stochastic noise and the signal has stronger self-organizing properties than noise, we may use NR-F to enhance its signal intensities and then improve its SNR. When the network has suffered from indirect noise that causes the false positive problem, we can treat the observed network as a diffusion from the real network and use NR-B to eliminate the indirect effect and then improve its SNR. We demonstrate the effectiveness of NR framework by applying it to several typical simulated and real-world biological and social networks. All the source code can be found at https://github.com/Wu-Lab/NR.

Although this paper has explored the denoising methods from two perspectives and demonstrates the effectiveness of the NR framework in different networks, network denoising is still an open problem to be solved because the properties of the different networks can be completely different, and it's unreasonable to improve the SNR of all noisy network using one single method. When the networks don't satisfy our assumptions, the denoising effect may not be ideal. Thus, it's valuable to explore the properties of network noise and signal comprehensively and propose the methods of network denoising from different angles. In addition, most network denoising algorithms need to transfer the developers' intuitions into rigorous mathematical languages for deeper understanding and interpretation of why and how they work, which is very important for promoting the research and application of network denoising and analysis in the future.

## Appendix

Appendix A contains the proofs of all the theorems in section II and Appendix B contains all the supplementary figures (Fig. S1−Fig. S12). Appendices have been given in separate files.

## APPENDIX A

Notice that the following three statements are equivalent to each other:
(1) The random walk on graph $G$ is reversible.
(2) The graph $G$ is undirected.
(3) The weighted adjacency matrix of graph $G$ is symmetric.

The operator $f_m (m > 1)$ is defined as:

$$f_m\colon\ \boldsymbol{\mathcal{P}} \longrightarrow \boldsymbol{\mathcal{P}}$$
$$f_m(P) = \frac{\sum_k \frac{P^k}{m^k}}{\sum_k \frac{1}{m^k}} = (m-1)P(mI-P)^{-1}$$

The operator $f_m^{-1}$ is defined as:

$$f_m^{-1}\colon\ \boldsymbol{\mathcal{P}} \longrightarrow \boldsymbol{\mathcal{P}}^*$$
$$f_m^{-1}(P) = m\big((m-1)I+P\big)^{-1}P$$

The operator $g$ is defined as:

$$g\colon\ \boldsymbol{\mathcal{W}} \rightarrow \boldsymbol{\mathcal{P}}$$
$$g(W) = D^{-1}W$$

The operator $h$ is defined as:

$$h: \boldsymbol{\mathcal{P}} \rightarrow \boldsymbol{\mathcal{W}}$$
$$h(P) = \alpha \cdot \mathrm{diag}(\pi(P))P$$

Now we give the proofs of Theorem 1—Theorem 7:

*Theorem 8.* $f_m(P) \in \boldsymbol{\mathcal{P}}$ for $P \in \boldsymbol{\mathcal{P}}$. That's to say:
(1) The sum of each row of the matrix $f_m(P)$ is 1.
(2) $f_m(P)$ is a non-negative matrix.

*Proof of Theorem 1:*

(1) Notice that the sum of each row of $P^l$ is 1, so the sum of each row of $P^l/m^l$ is $1/m^l$, hence the sum of each row of $f_m(P)$ is $\sum_l \frac{1}{m^l}/\sum_k \frac{1}{m^k} = 1$. Therefore $f_m(P)$ is a row stochastic matrix for $m > 1$.

(2) It's easy to verify using the series form of $f_m(P)$. ■

*Theorem 9.* Let $P$ be a transition matrix and $\{X_k\}$ be the random walk defined by $P$. The operator $f_m$ keeps the following properties of $P$ and $\{X_k\}$ unchanged:

(1) The reversibility of random walk $\{X_k\}$, thus the undirected property of the underlying graph.

(2) The stationary distribution of random walk $\{X_k\}$.

(3) Let $G$ be the graph where $\{X_k\}$ be treated as a random walk, then the degree distribution of $G$ keeps unchanged under operator $f_m$.

*Proof of Theorem 2:*

(1) If the input Markov chain is reversible, then there exists a reversible distribution $\pi$ satisfying $\pi_i P_{ij} = \pi_j P_{ji}$ for it. It's not difficult to get that $\pi_i (P^k)_{ij} = \pi_j (P^k)_{ji}$ for any positive integer $k$. In fact, when $k = 2$, we have:

$$\pi_i (P^2)_{ij} = \pi_i \sum_l P_{il}P_{lj} = \sum_l \pi_l P_{li} P_{lj} = \sum_l P_{li}\pi_j P_{jl}$$
$$= \pi_j \sum_l P_{jl}P_{li} = \pi_j (P^2)_{ji}$$

And we can get the conclusion recursively when $k > 2$. So $f_m(P)$ shares the same reversible distribution with $P$, thus the output Markov chain is reversible too.

(2) Let $\pi$ be the stationary distribution of $P$ such that $\pi P = \pi$, then we have $\pi P^k = \pi$, thus $\pi f_m(P) = \pi$ because $f_m(P)$ is a polynomial of $P$.

(3) We will prove that later in theorem 6. ■

*Theorem 10.* $f_m$ and $f_m^{-1}$ are inverse operators to each other.

*Proof of Theorem 3:*
(1) $f_m^{-1}\big(f_m(P)\big) = P$.
Denote $Y = f_m(P) = (m-1)P(mI-P)^{-1}$ for any transition matrix $P$. Then

$$Y(mI-P) = (m-1)P \quad \Rightarrow$$
$$mY = (m-1)P + YP = \big((m-1)I+Y\big)P \quad \Rightarrow$$
$$m\big((m-1)I+Y\big)^{-1}Y = P \quad \Rightarrow$$
$$f_m^{-1}(Y) = P$$

So we have $f_m^{-1}\big(f_m(P)\big) = P$.
(2) $f_m\big(f_m^{-1}(P)\big) = P$.
This can be proved in the same way as (1). ■

*Theorem 11.* Let $P$ be a transition matrix and $\{X_k\}$ be the random walk defined by $P$. The operator $f_m^{-1}$ keeps the following properties of $P$ $and$ $\{X_k\}$ unchanged:

(1) The reversibility of random walk $\{X_k\}$, thus the undirected property of the underlying graph.

(2) The stationary distribution of random walk $\{X_k\}$.

(3) Let $G$ be the graph where $\{X_k\}$ be treated as a random walk, then the degree distribution of $G$ keeps unchanged under operator $f_m^{-1}$.

*Proof of Theorem 4:*

(1) If the input Markov chain is reversible, then there exists a reversible distribution $\pi$ satisfying $\pi_i P_{ij} = \pi_j P_{ji}$ for it. Denote $\Pi = diag(\pi_1, \ldots, \pi_n)$, then we have $\Pi P = P^T \Pi$. Notice that $\Pi I = \Pi P P^{-1} = P^T \Pi P^{-1}$, so $\Pi P^{-1} = (P^T)^{-1}\Pi$. Then

$$(P^{-1})^T \Pi = \Pi P^{-1} \quad \Rightarrow$$
$$((m-1)P^{-1}+I)^T \Pi = \Pi\big((m-1)P^{-1}+I\big) \quad \Rightarrow$$
$$\Big(P^{-1}\big((m-1)I+P\big)\Big)^T \Pi = \Pi P^{-1}\big((m-1)I+P\big) \quad \Rightarrow$$

$$\Pi m\left((m-1)I+P\right)^{-1}P=\left(m\left((m-1)I+P\right)^{-1}P\right)^{T}\Pi$$

So we have $\Pi f_m^{-1}(P)=\left(f_m^{-1}(P)\right)^T\Pi$.

(2) Let $\pi$ be the stationary distribution of $P$ such that $\pi P=\pi$. Notice that $\pi I=\pi=\pi PP^{-1}=\pi P^{-1}$, so

$$\begin{aligned}\pi=\pi P^{-1} \quad &\Rightarrow\\ m\pi=(m-1)\pi P^{-1}+\pi=\pi P^{-1}\left((m-1)I+P\right) \quad &\Rightarrow\\ \pi m\left((m-1)I+P\right)^{-1}P=\pi\end{aligned}$$

which proves $\pi f_m^{-1}(P)=\pi$, so $\pi$ is the stationary distribution of $f_m^{-1}(P)$.

(3) We will prove that later in theorem 6. ■

*Theorem 12.* The sum of each row of the matrix $f_m^{-1}(P)$ is 1.

*Proof of Theorem 5:*

Denote $\mathbf{1}$ as the column vector whose elements are all 1. To prove the sum of each row of the matrix $f_m^{-1}(P)$ is 1, we only need to prove:

$$m\left((m-1)I+P\right)^{-1}P\mathbf{1}=\mathbf{1}$$

Namely, we need to prove:

$$m*\mathbf{1}=((m-1)I+P)\mathbf{1}$$

Note that the sum of each row of $I$ and $P$ are both 1, so the sum of each row of $(m-1)I+P$ is $m$. ■

*Theorem 13.* The operators $g$ and $h$ are inverse operators to each other.

*Proof of Theorem 6:*

(1) $h\left(g(W)\right)=W$.

According to the definition of stationary distribution, the stationary distribution $\pi$ of $D^{-1}W$ is $(\mathbf{1}^TD)/(\mathbf{1}^TD\mathbf{1})$ when $W$ is a symmetric matrix:

a) $(\mathbf{1}^TD)/(\mathbf{1}^TD\mathbf{1})\geq 0$ is obvious to see.

b) $(\mathbf{1}^TD)/(\mathbf{1}^TD\mathbf{1})\,\mathbf{1}=\mathbf{1}$.

c) $(\mathbf{1}^TD)/(\mathbf{1}^TD\mathbf{1})\,D^{-1}W=(\mathbf{1}^TW)/(\mathbf{1}^TD\mathbf{1}) = (\mathbf{1}^TD)/(\mathbf{1}^TD\mathbf{1})$.

So we have:

$$\begin{aligned}h\left(g(W)\right)&=h(D^{-1}W)=\alpha\mathrm{diag}\left(\frac{1^TD}{1^TD1}\right)D^{-1}W\\&=\mathrm{diag}(1^TD)D^{-1}W=W\end{aligned}$$

where $\alpha=\mathbf{1}^TD\mathbf{1}=\mathbf{1}^TW\mathbf{1}$, namely $\alpha$ is the total sum of matrix $W$.

(2) $g(h(P))=P$.

Assuming the stationary distribution of $P$ is $\pi(P)=(\pi_1,\ldots,\pi_n)$, then

$$h(P)=\alpha\begin{bmatrix}\pi_1 & & \\ & \ddots & \\ & & \pi_n\end{bmatrix}P$$

It's easy to check that the sum of each row of $h(P)$ is $(\alpha\pi_1,\ldots,\alpha\pi_n)$, so we have:

$$g\left(h(P)\right)=\begin{bmatrix}\frac{1}{\alpha\pi_1} & & \\ & \ddots & \\ & & \frac{1}{\alpha\pi_n}\end{bmatrix}\begin{bmatrix}\alpha\pi_1 & & \\ & \ddots & \\ & & \alpha\pi_n\end{bmatrix}P=P \quad ■$$

Now we treat back to the proof of Theorem 2 (3) and Theorem 2 (3). Because the degree matrix $D$ of the undirected graph can be computed by the stationary distribution of the random walk it defines, thus the degree distribution keeps unchanged if the stationary distribution keeps unchanged under operator $f_m$ and $f_m^{-1}$.

*Theorem 14.* For any input matrix $W$, the $m$-th order edge weight matrix of $F_m(W)$ defined by the $m$-th step transition matrix of $g(W)$ is (ignoring constant multiplies)：

$$\sum_{k_1,k_2,\ldots k_{m-1}}\left(\frac{W_{ik_1}W_{k_1k_2}\cdots W_{k_{m-1}j}}{\sum_l W_{k_1l}*\sum_l W_{k_2l}\cdots*\sum_l W_{k_{m-1}l}}\right)$$

which is a degree normalized version of the $m$-th order edge weight matrix of ND algorithm：

$$\sum_{k_1,k_2,\ldots k_{m-1}}\left(W_{ik_1}W_{k_1k_2}\cdots W_{k_{m-1}j}\right)$$

*Proof of Theorem 7:*

Denote $P=(P_{ij})$ as the transition matrix of input matrix $W$:

$$P_{ij}=\frac{W_{ij}}{\sum_l W_{il}}$$

Then the $m$-th step transition matrix of $W$ is:

$$\begin{aligned}\boldsymbol{P_{ij}^{(m)}}&=\sum_{k_1,k_2,\ldots k_{m-1}}P_{ik_1}P_{k_1k_2}\cdots P_{k_{m-1}j}\\&=\sum_{k_1,k_2,\ldots k_{m-1}}\left(\frac{W_{ik_1}}{\sum_l W_{il}}\frac{W_{k_1k_2}}{\sum_l W_{k_1l}}\cdots\frac{W_{k_{m-1}j}}{\sum_l W_{k_{m-1}l}}\right)\\&=\frac{1}{\sum_l W_{il}}\sum_{k_1,k_2,\ldots k_{m-1}}\left(\frac{W_{ik_1}W_{k_1k_2}\cdots W_{k_{m-1}j}}{\sum_l W_{k_1l}*\sum_l W_{k_2l}\cdots*\sum_l W_{k_{m-1}l}}\right)\end{aligned}$$

The $m$-th step weight matrix $W^{(m)} = \left(W_{ij}^{(m)}\right)$ of $W$ is:

$$W^{(m)} = h\left(P^{(m)}\right) = \alpha \cdot \mathrm{diag}\left(\pi\left(P^{(m)}\right)\right) P^{(m)}$$

According to (2) of theorem 2, the stationary distribution of random walk keeps unchanged under the operator $f_m$, so does the $m$-th step transition matrix, thus we have:

$$W^{(m)} = \alpha \cdot \mathrm{diag}\left(\pi(P)\right) P^{(m)}$$

Notice that the stationary distribution $\pi = (\pi_1, \dots, \pi_n)$ of $P = D^{-1}W$ is $(\mathbf{1}^T D)/(\mathbf{1}^T D \mathbf{1}) = (\mathbf{1}^T W)/\alpha$, then we have:

$$\begin{aligned} \boldsymbol{W_{ij}^{(m)}} &= \alpha \pi_i P_{ij}^{(m)} \\ &= \frac{\alpha \sum_l W_{il}}{\alpha \sum_l W_{il}} \sum_{k_1, k_2, \dots k_{m-1}} \left( \frac{W_{ik_1} W_{k_1 k_2} \cdots W_{k_{m-1} j}}{\sum_l W_{k_1 l} * \sum_l W_{k_2 l} \cdots * \sum_l W_{k_{m-1} l}} \right) \\ &= \sum_{k_1, k_2, \dots k_{m-1}} \left( \frac{W_{ik_1} W_{k_1 k_2} \cdots W_{k_{m-1} j}}{\sum_l W_{k_1 l} * \sum_l W_{k_2 l} \cdots * \sum_l W_{k_{m-1} l}} \right) \end{aligned}$$

## APPENDIX B

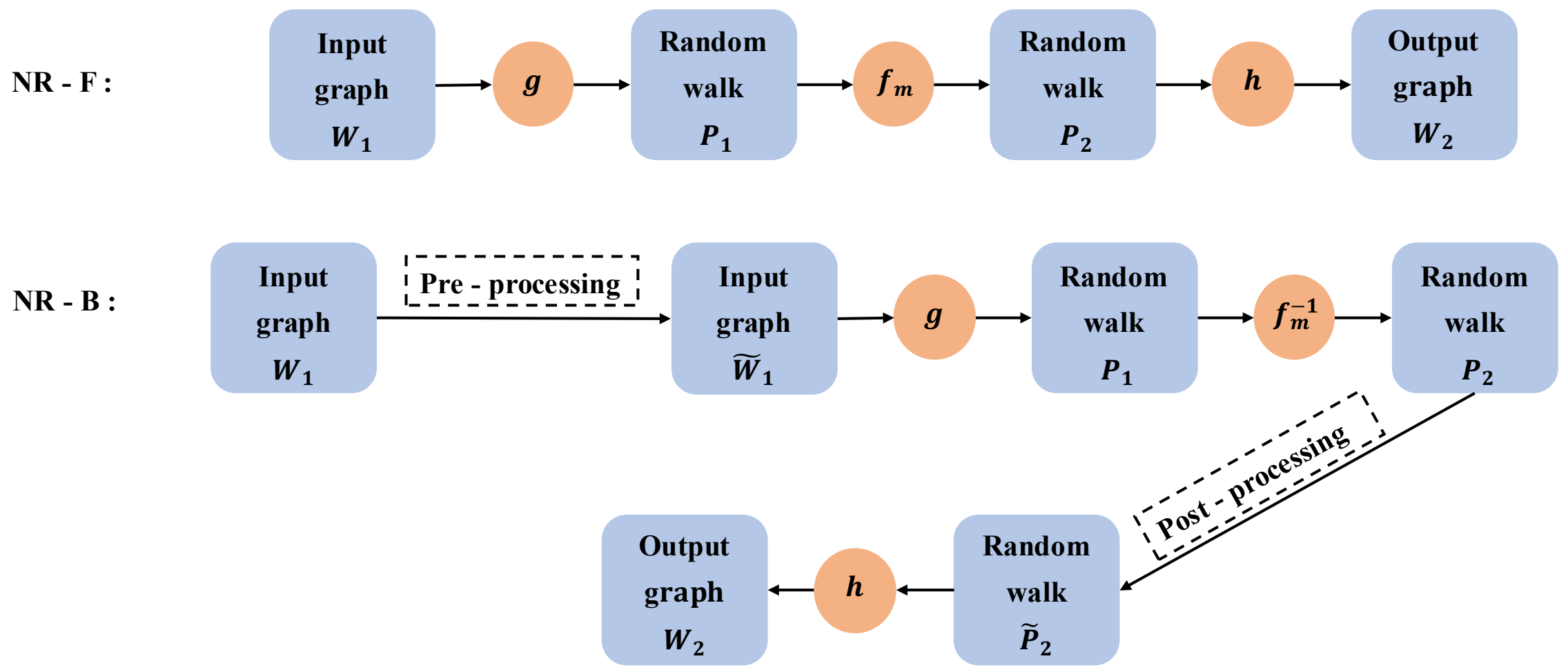

**S1. The detailed implementation flow of the NR framework.**

This figure corresponds to the simplified workflow (Figure 1c) in the main text. When applying NR-B, we may need to take pre-processing or post-processing steps in order to ensure that it runs smoothly.

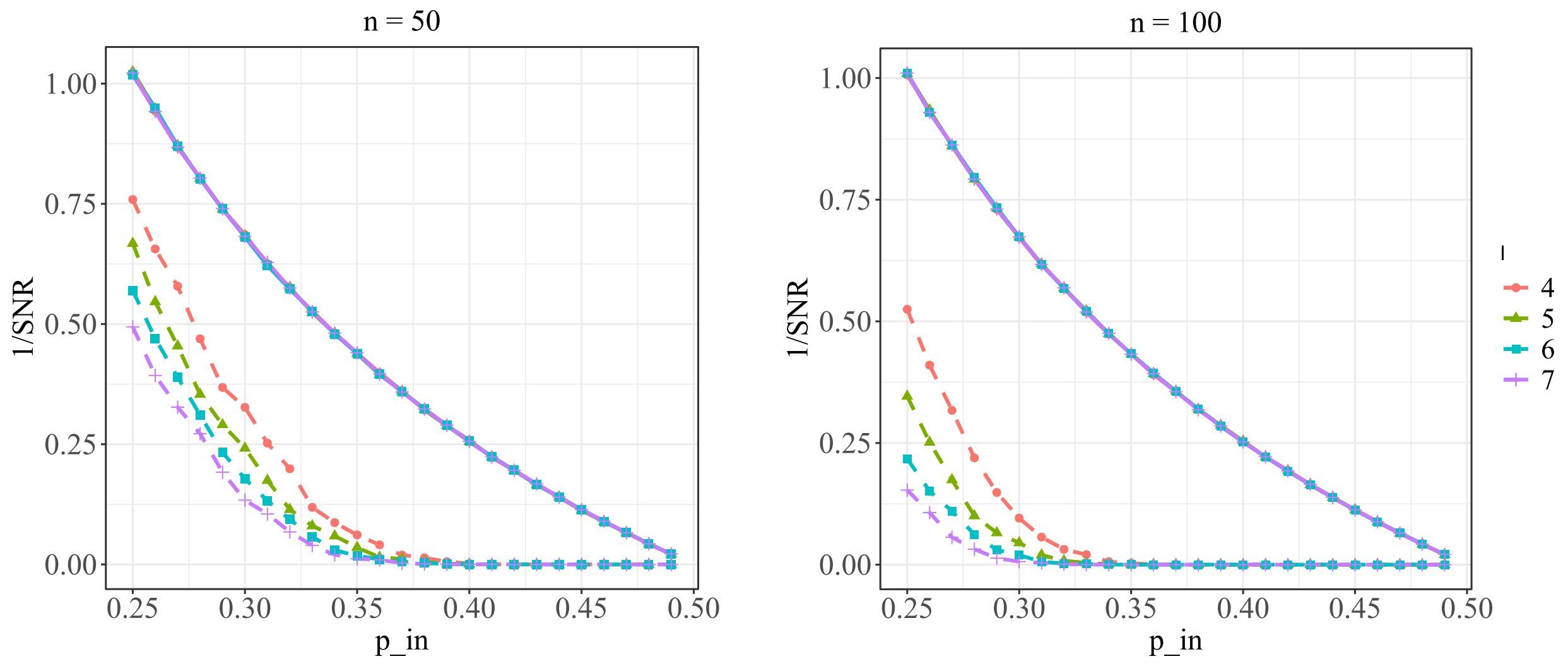

**S2. NR-F improves the SNR of simulated noisy networks.**

This figure shows the denoising performance of NR-F on the benchmark network given by Girvan and Newman under different network settings when $n = 50, l = 4,5,6,7$ and $n = 100, l = 4,5,6,7$. The dotted lines represent the NR-F denoised networks and the solid lines represent the original networks. Notice that the 1/SNR of the original network is the same for different values $l$, because the number of expected noise (edges between groups) and signal (edges inside groups) of the original network is $p_{out}C_l^2n^2$ and $p_{in}C_n^2l$ respectively, thus the 1/SNR is a function of $p_{in}$ and $n$, which is irrelevant to $l$. The results of each network configuration are the average of 50 experiments. The results demonstrate the denoising effectiveness of NR-F under different levels of denoising difficulty.

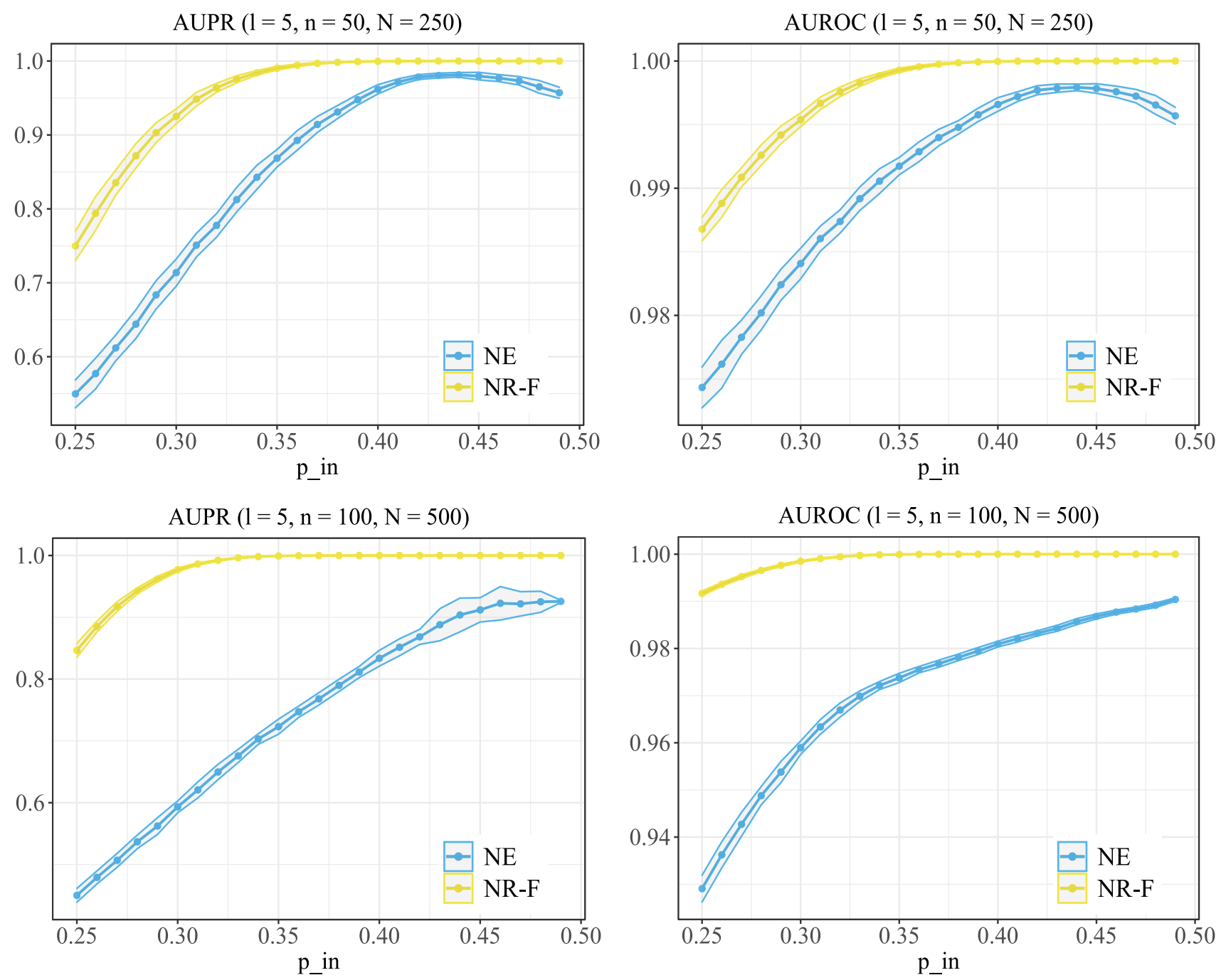

**S3. The AUPR and AUROC scores of NR-F and NE on simulated noisy networks.**

The figure shows the recovery accuracy of the signal edges of the benchmark network given by Girvan and Newman under different network settings: $N = 250, N = 500$. The denoising performance of NR-F and NE are shown in yellow and blue respectively. The results for each network configuration are the average of 50 experiments. It is clear that NR-F has a better denoising effect than NE under different evaluation metrics.

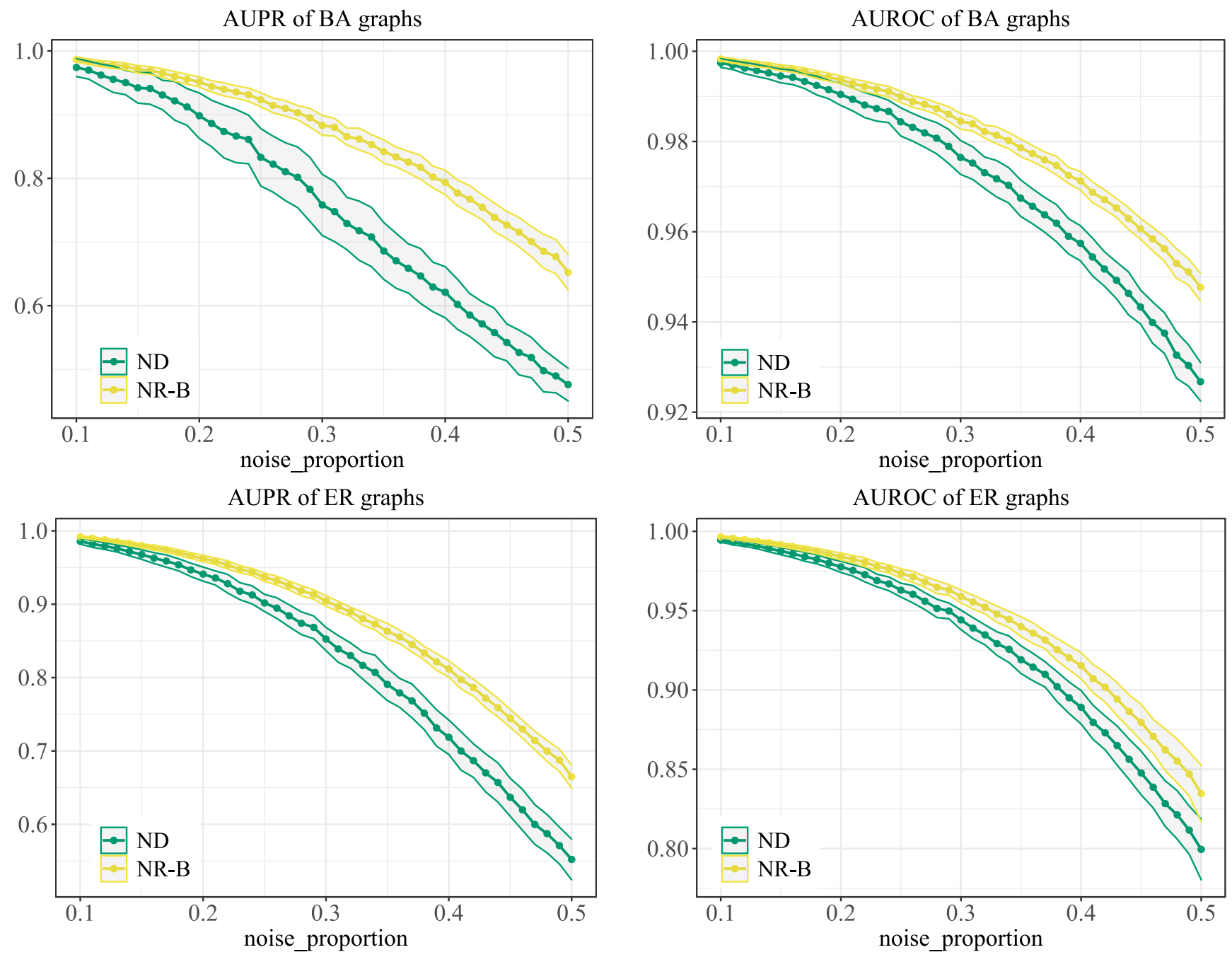

**S4. The AUPR and AUROC scores of NR-B and ND on simulated noisy networks.**

The figure shows the recovery accuracy of noisy ER graphs and noisy BA graphs. The original ER graph takes the parameter $N = 50, p = 0.3$ and the original BA graph takes the parameter $N = 100, m = 0.3$. The denoising performance of NR-B and ND are shown in yellow and green respectively. For each proportion of noisy edges, we show the mean score of 100 repeated tests. It is clear that NR-B has a better denoising effect than ND under different evaluation metrics.

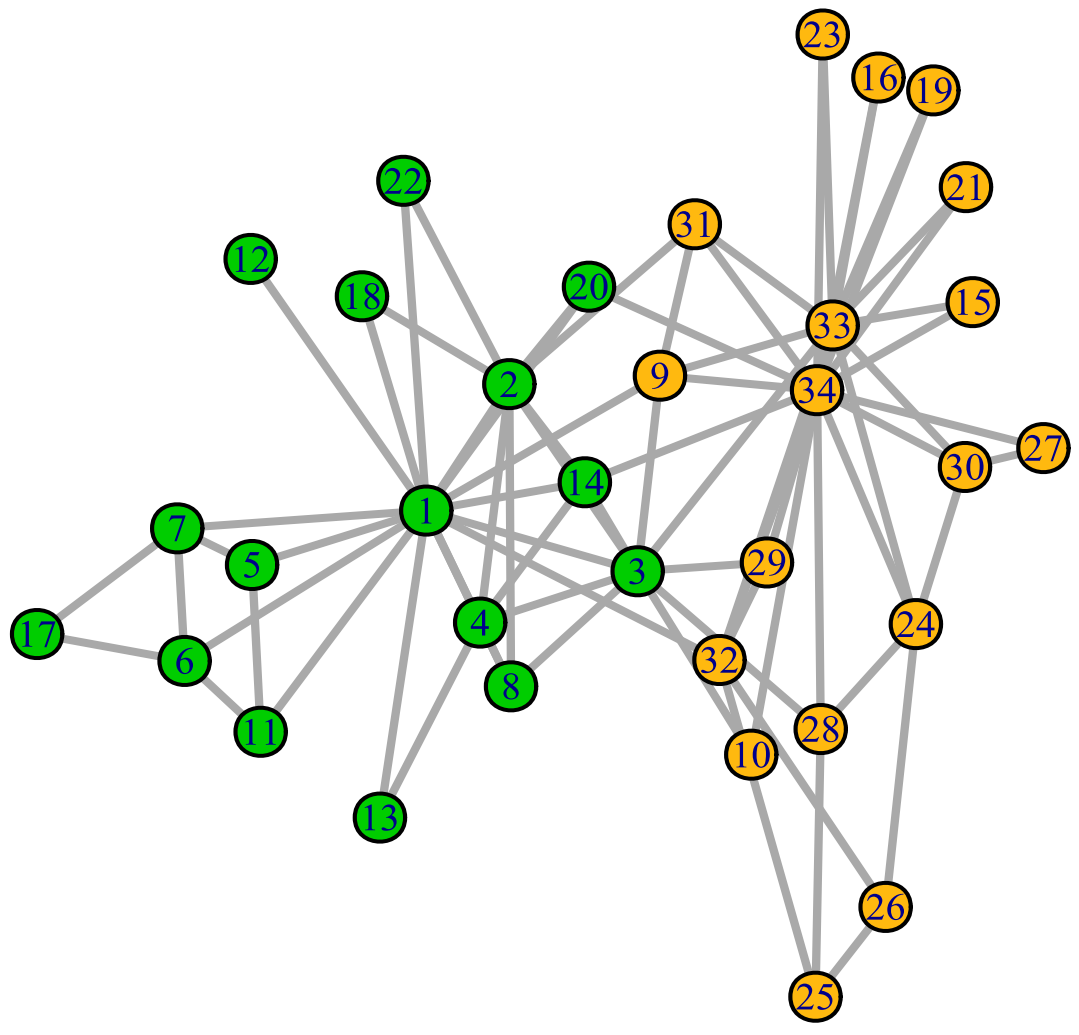

**S5. Zachary's karate club, a standard benchmark in community detection.**

It consists of 34 vertices representing the members of the club, the edge between vertices indicates the relationship between members. A conflict splits the club into two groups gathering around vertex 34 (the president) and 1 (the instructor).

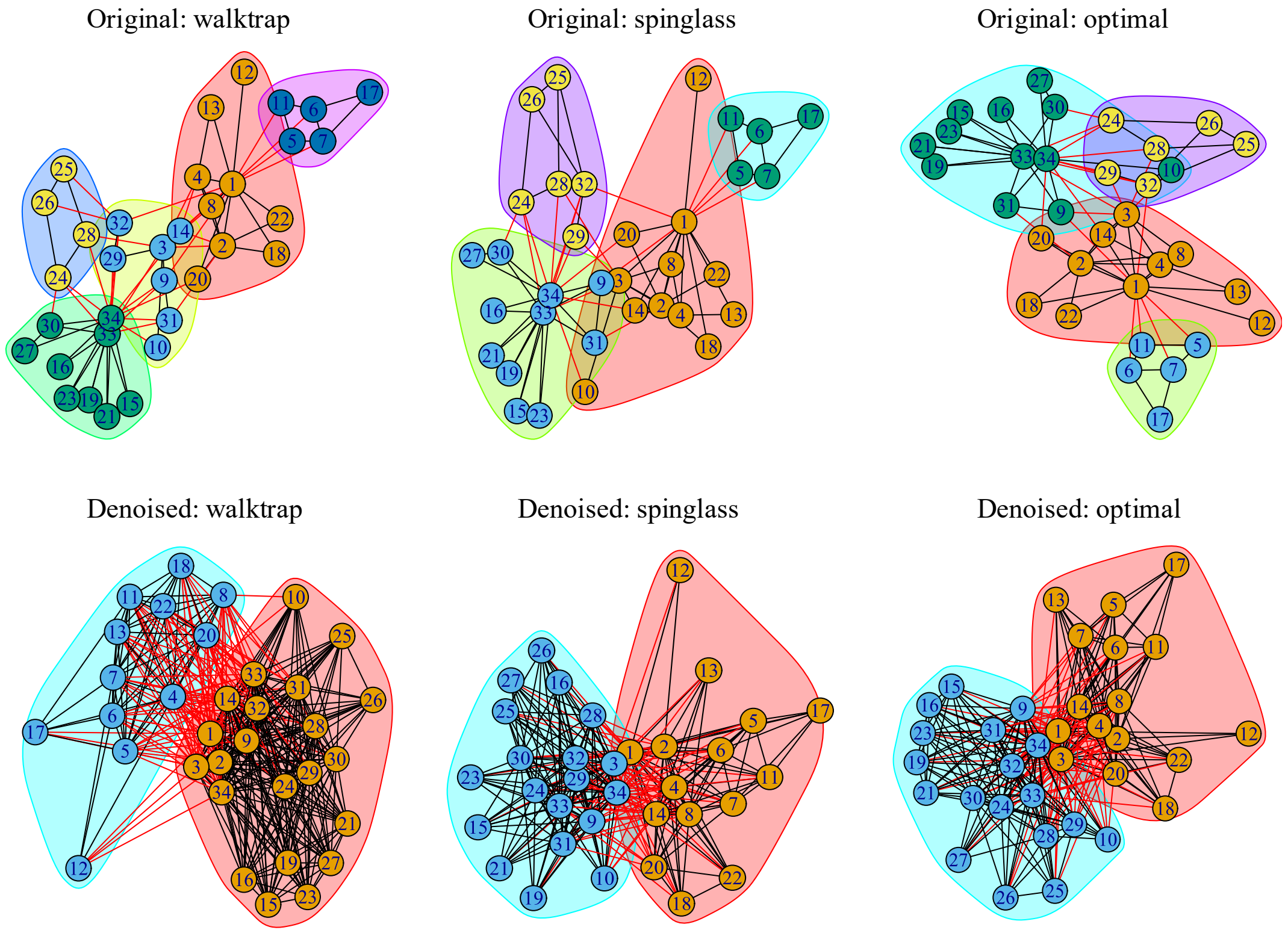

**S6. Demonstration of the effectiveness of NR-F in improving the accuracy of community detection on Zachary's karate club network.**

We compare the community detection results of walktrap, spinglass, and optimal algorithms on the network before and after applying NR-F as a denoising step. For display purposes, we set an appropriate threshold to change the weighted network into an unweighted network. This figure is a supplement to Figure 4a in the main text. It is worth noting that the spinglass algorithm gets different results every time, so we take the average of 20 trials.

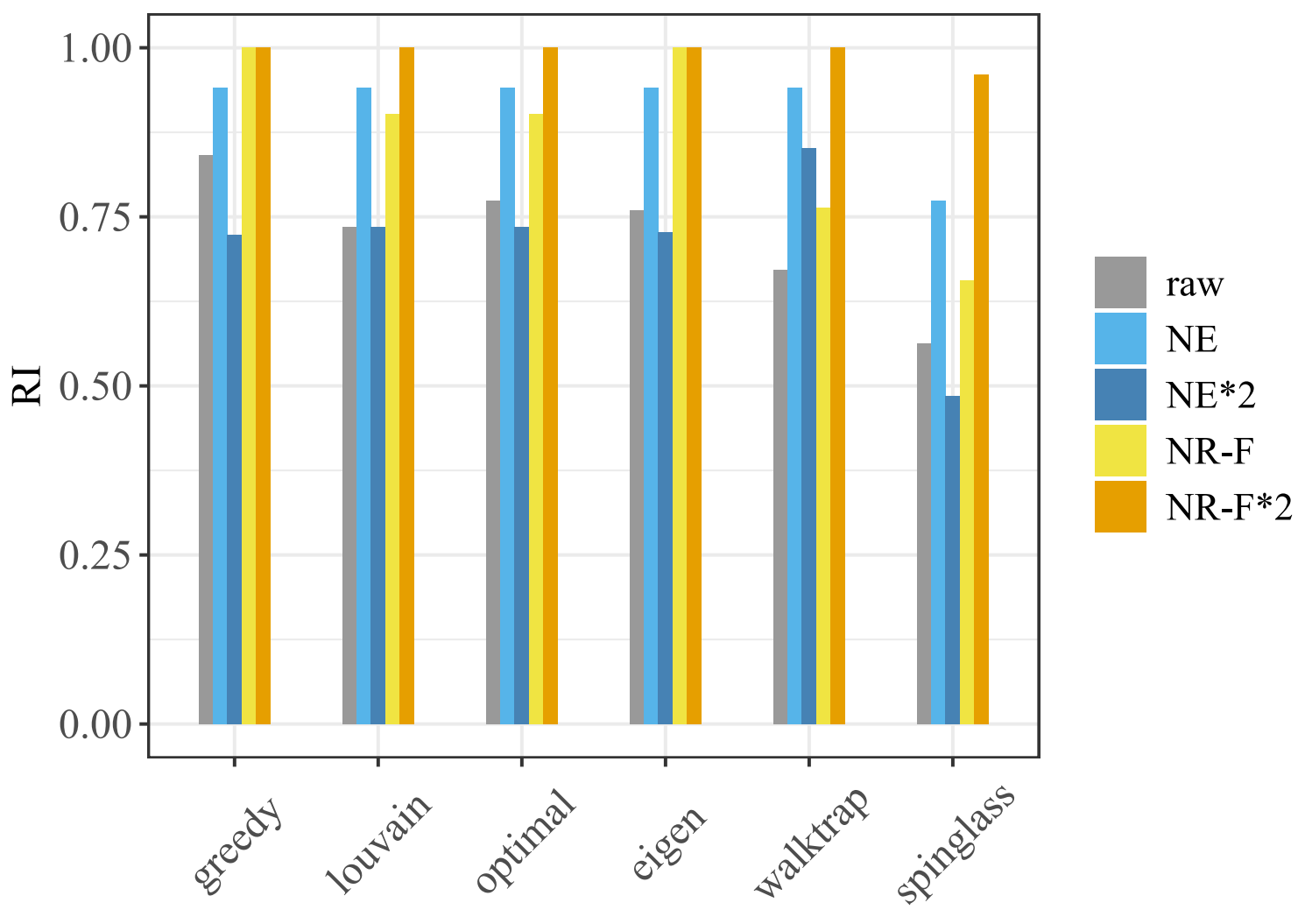

**S7. Improvement of the RI evaluation criteria on Zachary's karate club network after applying NR-F and NE as a denoising step.**

It is easy to see that after applying NR-F twice, basically, all community detection algorithms can classify correctly. The grey, blue, dark blue, yellow, orange bars represent the clustering performance on the original Zachary's karate network, the networks denoised by NE once and twice, the networks denoised by NR-F once and twice, respectively. This figure corresponds to Figure 4b in the main text.

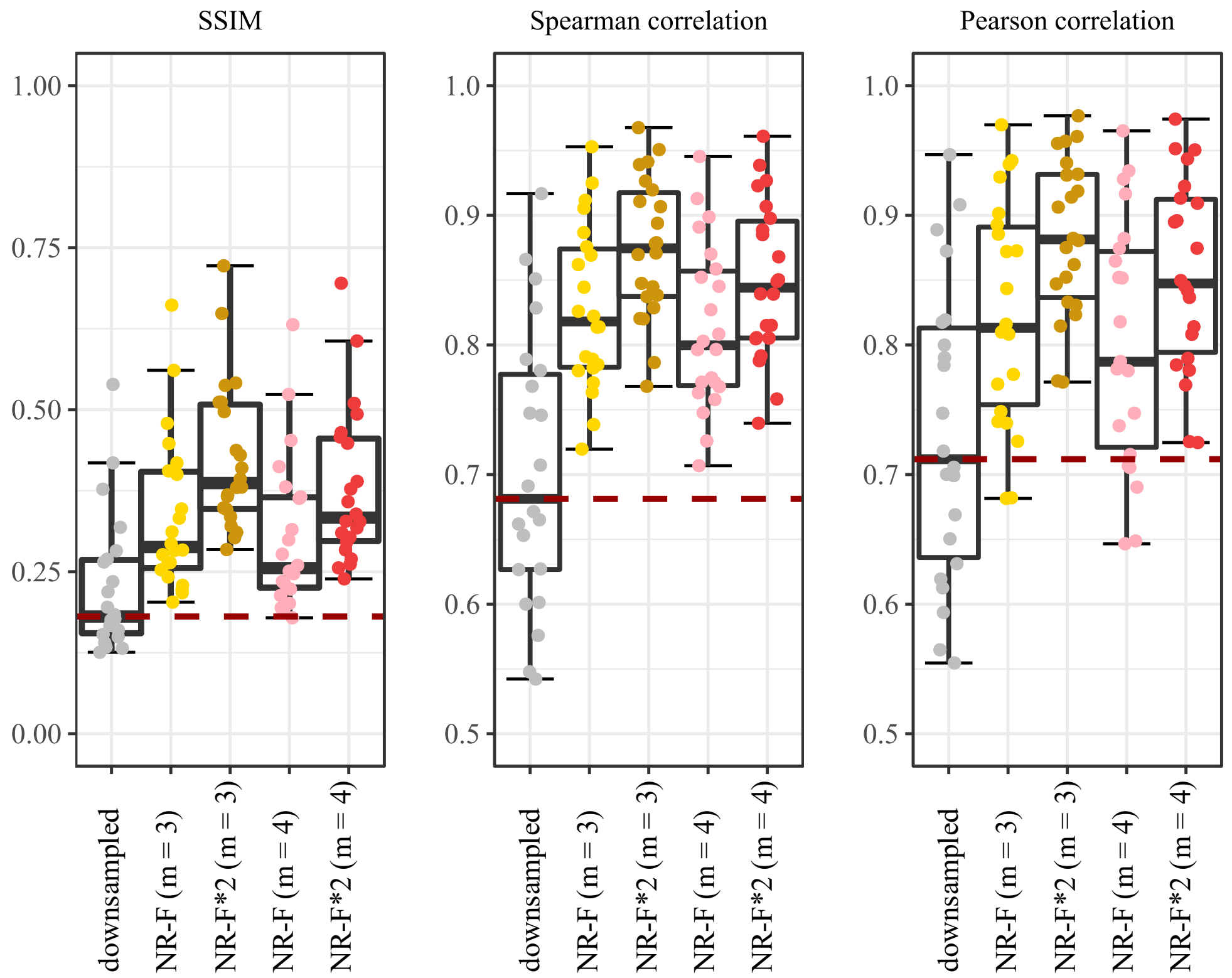

**S8. Parameter sensitivity analysis of NR-F in the experiment of enhancing the resolution of Hi-C data.**

The figure shows the denoising performance of applying NR-F with different parameters $m$ ($m = 3$ and $m = 4$) and different times (once and twice) under the evaluations of SSIM, Pearson correlation, and Spearman correlation. The results show that NR-F is insensitive to the values of parameter $m$ and application times since it has significant denoising effects in all cases.

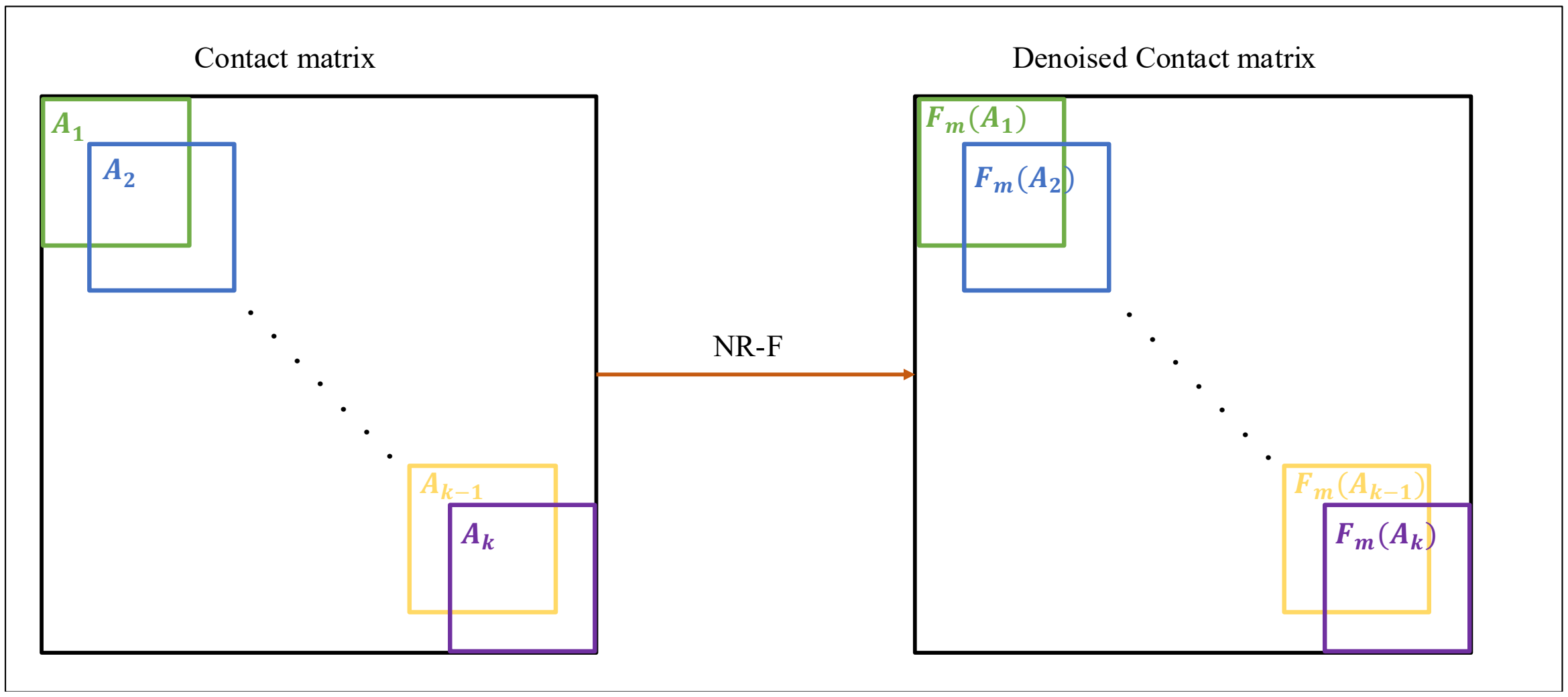

**S9. Illustration of the application of NR-F to the contact matrix in the experiment of enhancing the quality of Hi-C contact maps from the human genome.**

We apply NR-F to the downsampled contact matrix to denoise them. Instead of denoising the entire matrix directly, we only denoise the overlapping submatrices in the diagonal. If we get different values in different sub-matrices, we take their average.

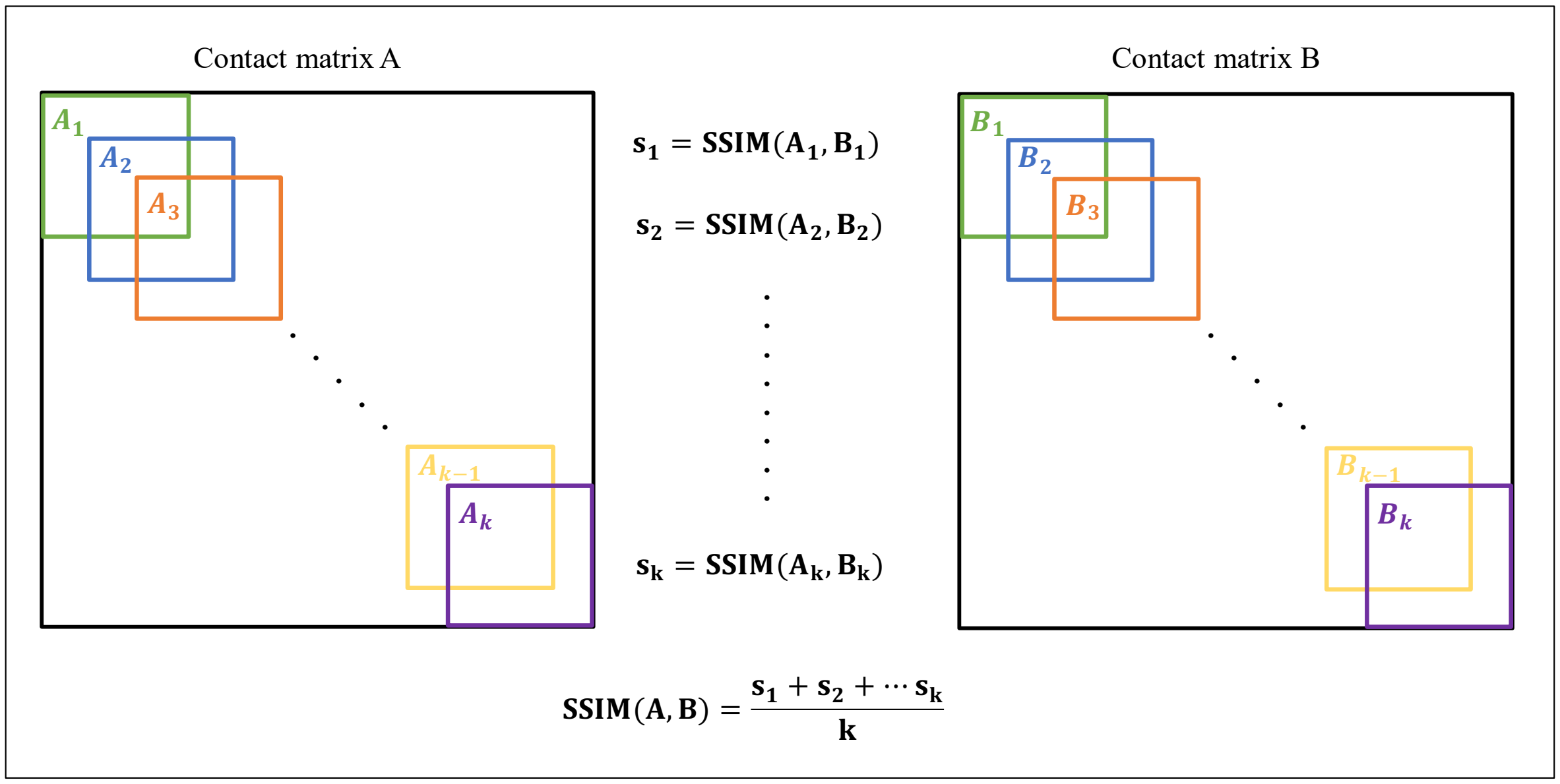

**S10. Calculation of SSIM evaluation criteria in the experiment of enhancing the quality of Hi-C contact maps from the human genome.**

The SSIM score between two contact matrixes is calculated by the average SSIM value of the overlapping sub-windows in the diagonal. Spearman correlation and Pearson correlation are calculated in the same way.

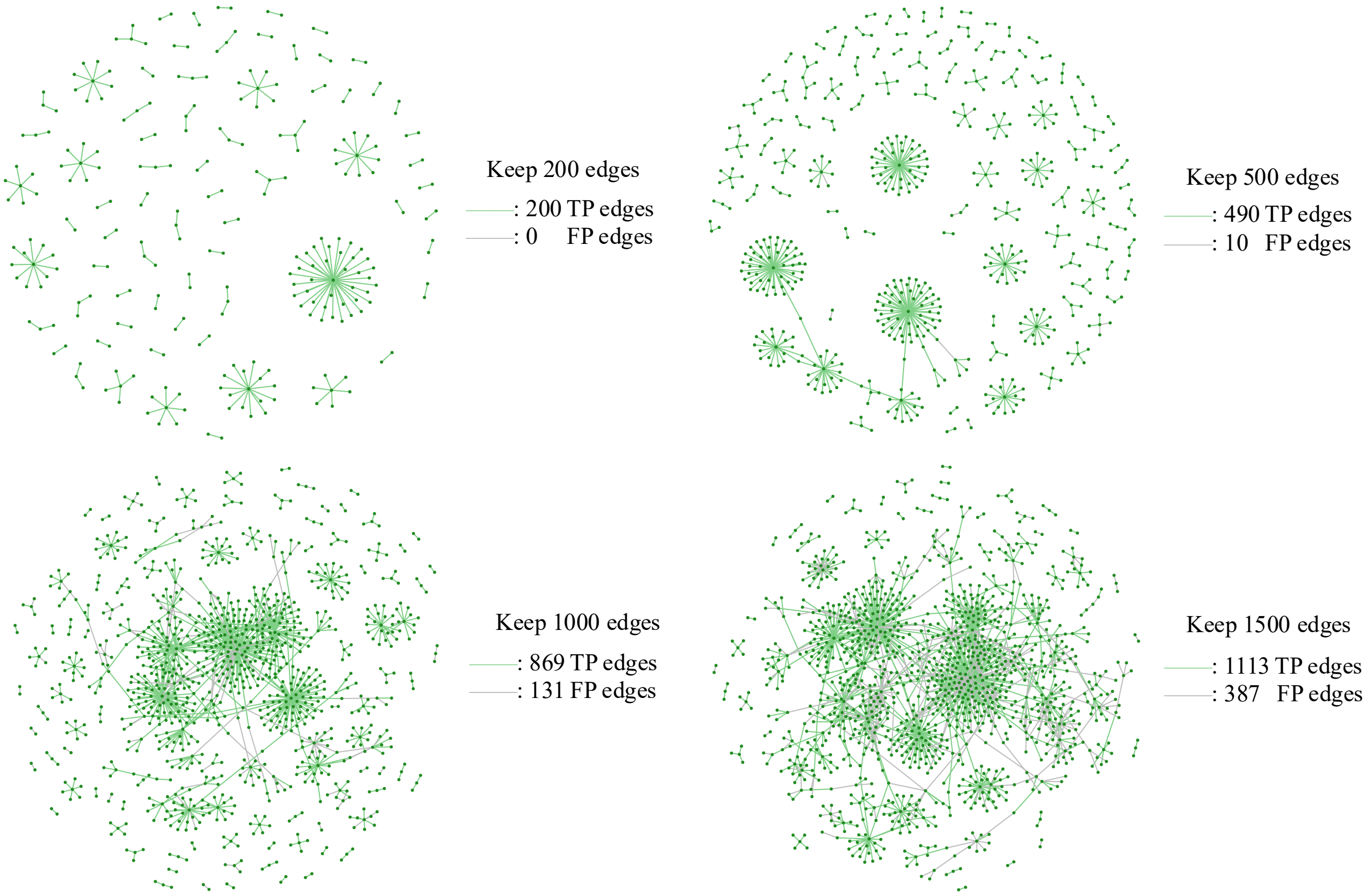

**S11. Presentation of the number of TP edges when different numbers of edges are retained on the NR-B denoised network.**

For the weighted network obtained by applying NR-B on the original weighted network inferred by GENIE3, we keep its top 200, 500, 1000, 1500 highest scoring edges and get 200, 490, 869, 1113 correctly inferred edges respectively.

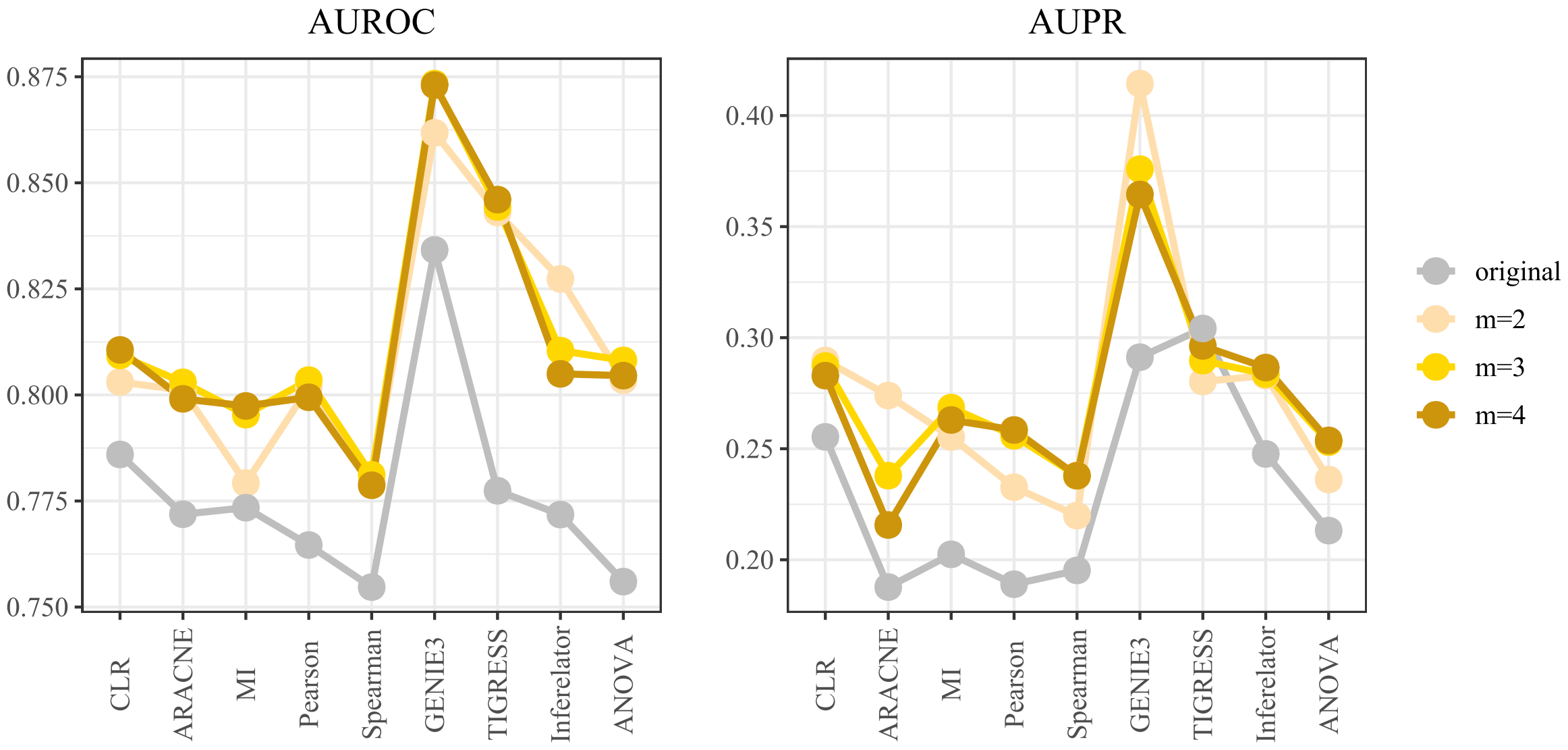

**S12. Parameter sensitivity analysis of NR-B in the experiment of refining the inference of gene regulatory networks.**

It can be seen that the different values of parameters $m$ have little effect on the results of the NR-B algorithm, which proves the robustness of NR-B to the parameter $m$.